\documentclass[twocolumn,twoside,10pt]{article}
\usepackage{graphicx,amsmath,amssymb,amsthm}

\makeatletter \oddsidemargin-.25in \evensidemargin-.25in
\makeatother \topmargin-0.75in \textwidth7in \textheight9.5in

\theoremstyle{definition}

\begin{document}

\date{}

% Title----------------------------------------------------------------
\title{\begin{flushleft}
\noindent {\small {\it submitted to Journal of Nonlinear Systems and Applications (Oct 2009)}}
\end{flushleft}\Large\bf \uppercase{Transfer Entropy on Rank Vectors}}
%\author{\Large\bf Author Index}
\author{Dimitris Kugiumtzis
 \thanks{D. Kugiumtzis is with the Department
of Mathematical, Physical and Computational Sciences, Faculty of Engineering, Aristotle
University of Thessaloniki, Thessaloniki 54124, Greece, e-mail: dkugiu@gen.auth.gr}}
 \maketitle

% Abstract----------------------------------------------------------------
{\footnotesize \noindent {\bf Abstract.}
Transfer entropy (TE) is a popular measure of information flow found to
perform consistently well in different settings. Symbolic transfer entropy (STE)
is defined similarly to TE but on the ranks of the components
of the reconstructed vectors rather than the reconstructed vectors themselves.
First, we correct STE by forming the ranks for the future samples
of the response system with regard to the current reconstructed vector.
We give the grounds for this modified version of STE, which we call Transfer Entropy on
Rank Vectors (TERV).
Then we propose to use more than one step ahead in the formation of the future
of the response in order to capture the information flow from the driving system
over a longer time horizon. To assess the performance of STE, TE and TERV in detecting correctly
the information flow we use receiver operating characteristic (ROC) curves
formed by the measure values in the two coupling directions computed on a number
of realizations of known weakly coupled systems. We also consider different
settings of state space reconstruction, time series length and
observational noise. The results show that TERV indeed improves STE and in some
cases performs better than TE, particularly in the presence of noise,
but overall TE gives more consistent results.
The use of multiple steps ahead improves the accuracy of TE and TERV.
 }

\vskip.2in

%%%%%%%%%%%%%%%%%%%%%%
\section{Introduction}
%%%%%%%%%%%%%%%%%%%%%%

The fundamental concept for the dependence of one variable $Y$ measured over
time on another variable $X$ measured synchronously is the Granger causality
\cite{Granger69}. While Granger defined the direction of interaction
in terms of the contribution of $X$ in predicting $Y$, many variations
of this concept have been developed, starting with
linear approaches in the time and frequency domain (e.g. see
\cite{Baccala01,Winterhalder05}) and extending
to nonlinear approaches focusing on phase or event synchronization
\cite{Rosenblum01,QuianQuiroga02,Smirnov03}, comparing neighborhoods of
the reconstructed points from the two time series
\cite{Cenys92,Schiff96,Arnhold99,QuianQuiroga00b,Andrzejak03b,
Romano07,Chicharro09}, and measuring the information
flow between the time series
\cite{Schreiber00,Palus01,Marschinski02,Staniek08,Vejmelka08}.

Among the different proposed measures we concentrate here on the last
class of measures, and particularly on the transfer entropy (TE)
\cite{Schreiber00} and the most recent variant of TE operating on
rank vectors, called symbolic transfer entropy (STE) \cite{Staniek08} (see
also \cite{Bahraminasab08} for a similar measure).
There have been a number of
comparative studies on information flow measures and other coupling
measures giving varying results. In all the studies where TE was considered,
it performed at least as good as the other measures
\cite{Lungarella07,Palus07,Papana08b}. The STE measure
is proposed as an improvement of TE in real world applications, where
noise may mask details of the fine structure, that can be better
treated by coarse discretization using ranks instead of samples.

We propose here a correction of STE. In the definition of TE
the observable of the response at one time step ahead is a scalar, but in STE
it is taken as a rank vector at this time index. We modify STE to conform
with the definition of TE and give grounds for the correctness of this
modification. Further, we allow for the future of the response to be
defined over more than one time steps. To the best of our knowledge this
has not been implemented in TE, but it is the core element in the information
flow measures of mean conditional mutual information \cite{Vejmelka08} and
coarse-grained transinformation rate \cite{Palus01}.
In many applications on interacting flow systems, the sampling time may be
small and a single step ahead may not regard the time of the response to
the directed coupling, as in electroencephalography (EEG)
\cite{Staniek08,Sabesan09b}, and in the analysis of financial indices
\cite{Marschinski02,Kwon08}. The same may hold for maps: the transfer of
information may better be seen over more than one iteration of the
interacting maps. We compare TE, STE and our correction of STE on measuring
weak directed interaction in some known coupled systems and for different state space
reconstructions, time series lengths and also in the presence of noise. We also
investigate the change in the performance of these measures when defining
them for more than one step ahead.

In the following, TE and STE measures are presented briefly in Section~\ref{sec:methods},
and the proposed modification of TE is described in Section~\ref{sec:proposed}.
Then the results of the simulation study comparing the proposed measure to TE and STE are presented in Section~\ref{sec:simulations}, and discussed in Section~\ref{sec:discussion}.

%%%%%%%%%%%%%%%%%%%%%%%%%%%%%%
\section{Information flow measures}
\label{sec:methods}
%%%%%%%%%%%%%%%%%%%%%%%%%%%%%%

Let us suppose that a representative quantity of system $X$ is measured giving
a scalar time series $\{x_t\}_{t=1}^N$ and that we have respectively $\{y_t\}_{t=1}^N$ for $Y$,
where $X$ and $Y$ possibly interact.
Using the method of delays the reconstructed points from the two time series are
$\mathbf{x}_t = [x_t,x_{t-\tau_x},\ldots,x_{t-(m_x-1)\tau_x}]$
and $\mathbf{y}_t = [y_t,y_{t-\tau_y},\ldots,y_{t-(m_y-1)\tau_y}]$, allowing
different delay parameters $\tau_x$, $\tau_y$ and embedding dimensions $m_x$, $m_y$
for the systems $X$ and $Y$, respectively.

{\em Transfer entropy} (TE) is a measure of the information flow from the
driving system to the response system. Specifically, TE estimates
the entropy in the response system
caused by its connection to the driving system, accounting for the entropy generated
internally in the response system \cite{Schreiber00}. TE for the causal effect of
system $X$ on system $Y$ can be defined in terms of the Shannon entropy
$H(x) = \sum p(x)\log p(x)$ as
%----------------------------------------------------------------------
\begin{eqnarray}
& & \mbox{TE}_{X\rightarrow Y} = \label{eq:TEH}
 \\
& & -H(y_{t+1},\mathbf{x}_t,\mathbf{y}_t) + H(\mathbf{x}_t,\mathbf{y}_t) + H(y_{t+1},\mathbf{y}_t) -H(\mathbf{y}_t),
\nonumber
\end{eqnarray}
%----------------------------------------------------------------------
or directly in terms of distribution functions as
%----------------------------------------------------------------------
\begin{equation}
\mbox{TE}_{X\rightarrow Y} = \sum p(y_{t+1},\mathbf{x}_t,\mathbf{y}_t)%
\log{\frac{p(y_{t+1}|\mathbf{x}_t,\mathbf{y}_t)}                      %
{p(y_{t+1}|\mathbf{y}_t)}},                                           %
\label{eq:TE}
\end{equation}
%----------------------------------------------------------------------
where $p(y_{t+1},\mathbf{x}_t,\mathbf{y}_t)$,
$p(y_{t+1}|\mathbf{x}_t,\mathbf{y}_t)$, and
$p(y_{t+1}|\mathbf{y}_t)$ are the joint and conditional
probability mass functions (pmf). The summation is over all the cells of
a suitable partition of the joint variable vectors appearing as arguments
in the pmfs or entropy terms.

The estimation of TE requires the estimation of the pmfs in eq.(\ref{eq:TE}),
or the probability density functions assuming the integral form and no binning.
The pmfs are estimated directly by the relative frequency of occurrence of
points in each cell, but finding a suitable binning may be challenging
\cite{Cover91,Papana09}. Moreover, for high-dimensional reconstructions, the binning
estimators are data demanding.
Therefore estimators of the probability density functions are more
appropriate for TE estimation, such as kernels \cite{Silverman86},
nearest neighbors \cite{Kraskov04}, and correlation sums \cite{Diks02}.
We follow the latter approach to estimate TE and recall first that without
assuming discretization each term of the form $H(\mathbf{x})$ in (\ref{eq:TEH})
expresses the differential entropy of the vector variable $\mathbf{x}$. The
differential entropy can be approximated from the correlation sum
$C(\mathbf{x})$ as $H(\mathbf{x}) \simeq \ln{C(\mathbf{x})} + m \ln{r}$,
where $C(\textbf{x})$ is the estimated cumulative density of
inter-point distances at embedding dimension $m$ and for a suitably small
distance $r$ \cite{Diks02}. Thus TE is
estimated by the correlation sums as
% -------------------------------------------------------------
\begin{equation}
\mbox{TE}_{X \rightarrow Y} =
\log{\frac{C(y_{t+1},\mathbf{x}_t,\mathbf{y}_t)C(\mathbf{y}_t)}
{C(\mathbf{x}_t,\mathbf{y}_t)C(y_{t+1},\mathbf{y}_t)}},
\label{eq:TECorSum}
\end{equation}
% -------------------------------------------------------------
where
$C(y_{t+1},\mathbf{x}_t,\mathbf{y}_t)$, $C(\mathbf{y}_t)$,
$C(\mathbf{x}_t,\mathbf{y}_t)$ and $C(y_{t+1},\mathbf{y}_t)$ are
the correlation sums for the points of the form
$[y_{t+1},\mathbf{x}_t,\mathbf{y}_t]$, $\mathbf{y}_t$,
$[\mathbf{x}_t,\mathbf{y}_t]$ and $[y_{t+1},\mathbf{y}_t]$,
respectively. The corresponding vector dimensions are $1+m_x+m_y, m_y,
m_x+m_y$ and $1+m_y$. To account for the different dimensions,
we use the standardized Euclidean norm for the distances.

The so-called {\em symbolic transfer entropy}
(STE) is derived as the transfer entropy defined on rank vectors formed by the
reconstructed points \cite{Staniek08}. For each point $\mathbf{y}_t$, the
ranks of its components in ascending order assign a rank vector $\hat{\mathbf{y}}_t =
[r_1,r_2,\ldots,r_{m_y}]$, where $r_j\in\{1,2,\ldots,m_y\}$ for
$j=1,\ldots,m_y$, is the rank order of the component $y_{t-(j-1)\tau_y}$
(for two equal components of $\mathbf{y}_t$ the smallest rank
is assigned to the component appearing first in $\mathbf{y}_t$).
Substituting also $y_{t+1}$ in eq.(\ref{eq:TEH}) with the rank vector at time
$t+1$, $\hat{\mathbf{y}}_{t+1}$, STE is defined as
%----------------------------------------------------------------------
\begin{eqnarray}
& & \mbox{STE}_{X\rightarrow Y} = \label{eq:STEH} \\
& & -H(\hat{\mathbf{y}}_{t+1},\hat{\mathbf{x}}_t,\hat{\mathbf{y}}_t) + H(\hat{\mathbf{x}}_t,\hat{\mathbf{y}}_t) + H(\hat{\mathbf{y}}_{t+1},\hat{\mathbf{y}}_t) -H(\hat{\mathbf{y}}_t). \nonumber
\end{eqnarray}
%----------------------------------------------------------------------
The estimation of STE from eq.(\ref{eq:STEH}) is straightforward as the
pmfs are naturally defined on the rank vectors.
There is a great advantage of using a rank vector
$\hat{\mathbf{y}}_t$ over a binning of $\mathbf{y}_t$, say using $b$ bins for each component: the possible vectors from binning are $b^{m_y}$ while
the possible combinations of the rank vectors are $m_y!$. For example,
for $b=m_y=4$, there are 256 cells from binning and only 24 combinations
of rank vectors. Still, the estimation of the probability of
occurrence of a rank vector becomes unstable as the dimension increases.
Especially, for the joint vector of ranks
$[\hat{\mathbf{y}}_{t+1},\hat{\mathbf{x}}_t,\hat{\mathbf{y}}_t]$ the dimension
is $2m_y+m_x$, for which the equivalent of TE is $[y_{t+1},\mathbf{x}_t,\mathbf{y}_t]$
and has dimension $1+m_x+m_y$.

%%%%%%%%%%%%%%%%%%%%%%%%%%%%%%%%%%%%%%%%%%%%%%%%%%
\section{Modification of symbolic transfer entropy}
\label{sec:proposed}
%%%%%%%%%%%%%%%%%%%%%%%%%%%%%%%%%%%%%%%%%%%%%%%%%%
The conversion of the scalar $y_{t+1}$ to the rank vector $\hat{\mathbf{y}}_{t+1}$
seems to have been chosen in order to express $y_{t+1}$ in terms of ranks in \cite{Staniek08}.
Under this conversion, STE is not the direct analogue
to TE using ranks instead of samples.
The problem is not so much the use of the scalar $y_{t+1}$ or the vector $\mathbf{y}_{t+1}$ in the
definition of TE in eq.(\ref{eq:TEH}) or eq.(\ref{eq:TE}) because for $\tau_y=1$
$p(y_{t+1},\mathbf{x}_t,\mathbf{y}_t)=p(\mathbf{y}_{t+1},\mathbf{x}_t,\mathbf{y}_t)$,
as all components but $y_{t+1}$ of the vector $\mathbf{y}_{t+1}$ are also
components of $\mathbf{y}_t$. The same holds for the conditional
pmfs in eq.(\ref{eq:TE}) and the two correlation sums in which $y_{t+1}$ appears
in eq.(\ref{eq:TECorSum}). We elaborate on the implication of the use of $\hat{\mathbf{y}}_{t+1}$
below.

\begin{figure}[htb!]
\centering
\includegraphics[height=3cm]{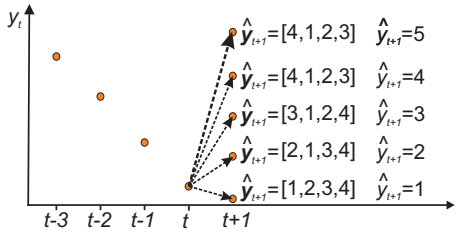}
\caption{Sketch of a position of samples $y_{t-3},y_{t-2},y_{t-1},y_{t}$ and the possible rank position of $y_{t+1}$ together with the corresponding rank vector $\hat{\mathbf{y}}_{t+1}$ defined for STE and the actual rank of $y_{t+1}$ considering all 5 samples.}
 \label{fig:sketchranks}
\end{figure}

Let us first assume that $\tau_y=1$. A first problem lies in the fact that when deriving the rank vector $\hat{\mathbf{y}}_{t+1}$ associated with $\mathbf{y}_{t+1}$, the  rank of the last component of $\mathbf{y}_t$, $y_{t-m_y+1}$, is not considered. As an example, consider
the vector $\mathbf{y}_t=[y_t,y_{t-1},y_{t-2},y_{t-3}]^{\prime}$ with a corresponding
rank vector $\hat{\mathbf{y}}_t=[1,2,3,4]$, i.e. the samples decrease with time. If the
decrease continues at the next time step then
$\hat{\mathbf{y}}_{t+1}=[1,2,3,4]$, if $y_{t+1}$ is between $y_t$ and $y_{t-1}$ then $\hat{\mathbf{y}}_{t+1}=[2,1,3,4]$,
if it is between $y_{t-1}$ and $y_{t-2}$ then  $\hat{\mathbf{y}}_{t+1}=[3,1,2,4]$, and finally if $y_{t+1}$ is
larger than $y_{t-2}$ (the largest of all components in $\mathbf{y}_{t+1}$) then $\hat{\mathbf{y}}_{t+1}=[4,1,2,3]$.
The 4 possible scenarios are shown in Fig.~\ref{fig:sketchranks}.
The definition of rank vector $\hat{\mathbf{y}}_{t+1}$ accounts only for the possible rank positions of $y_{t+1}$ with respect to the last $m_y-1$ samples, ignoring the sample $y_{t-m_y+1}$, here $y_{t-3}$. With regard to the same example, $\hat{\mathbf{y}}_{t+1}=[4,1,2,3]$ assigns to both cases $y_{t-2} < y_{t+1} < y_{t-3}$ and $y_{t-3} < y_{t+1}$ (see Fig.~\ref{fig:sketchranks}).
In the entropy or probability terms of the definition of TE, $y_{t+1}$ appears together with $\mathbf{y}_t$, and there are 5 possible rank positions of $y_{t+1}$ in the augmented vector $[y_{t+1},y_t,y_{t-1},y_{t-2},y_{t-3}]$, as shown in Fig.~\ref{fig:sketchranks}. Thus for $m_y=4$ there are $5!=120$ different rank orders for the joint vector
$[y_{t+1},\mathbf{y}_t]$, but when forming the joint rank vector $[\hat{\mathbf{y}}_{t+1},\hat{\mathbf{y}}_t]$ (as in the computation of STE) there are
only $4!\cdot(4!/3!)=96$ possible rank orders. In general, there are $(m_y+1)!$ possible rank orders for the joint vector $[y_{t+1},\mathbf{y}_t]$, but STE estimation represents them in $m_y!\cdot\frac{m_y!}{(m_y-1)!}$ rank orders of $[\hat{\mathbf{y}}_{t+1},\hat{\mathbf{y}}_t]$.

\begin{figure}[htb!]
\centering
\hbox{\includegraphics[height=3.5cm]{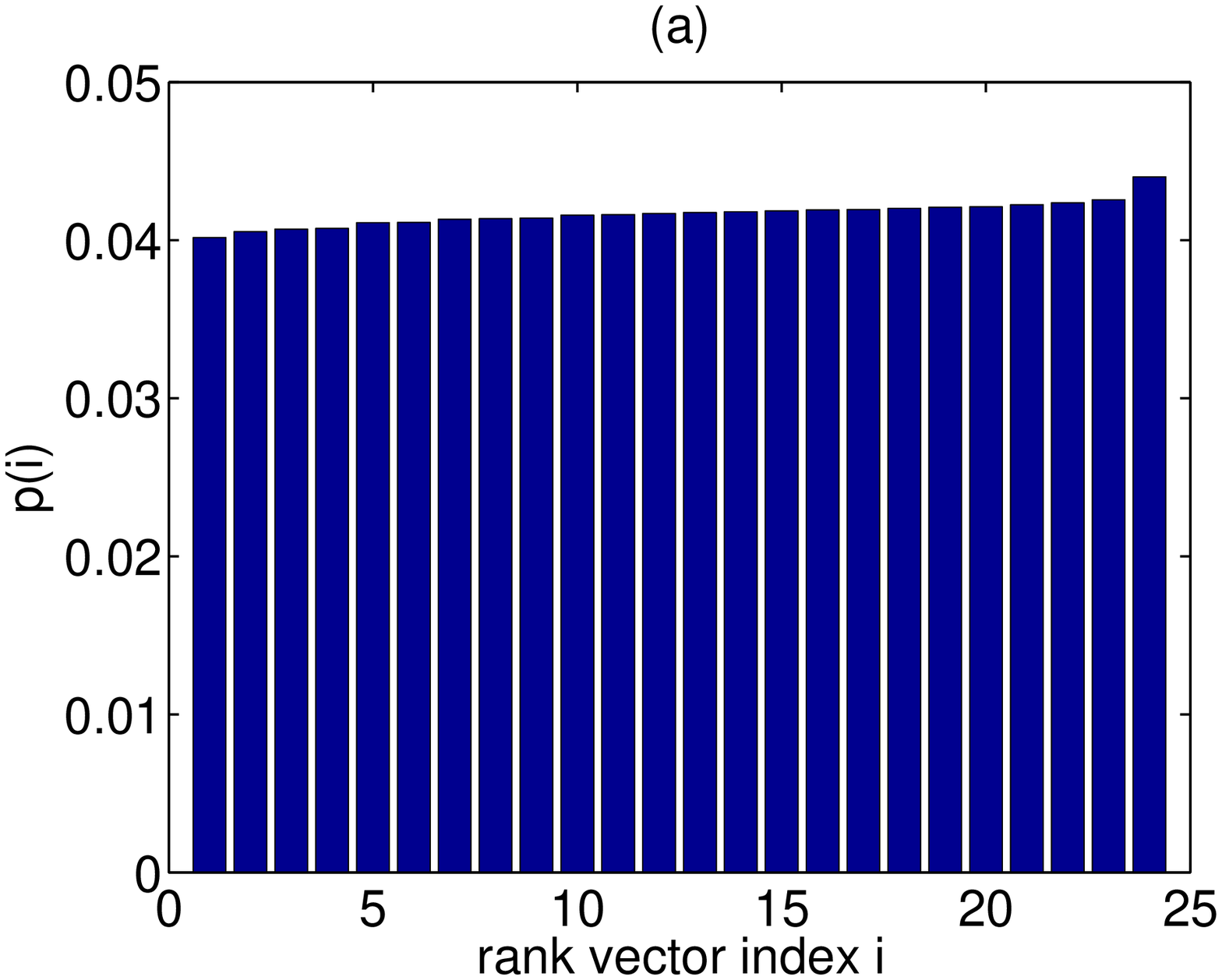}
\includegraphics[height=3.5cm]{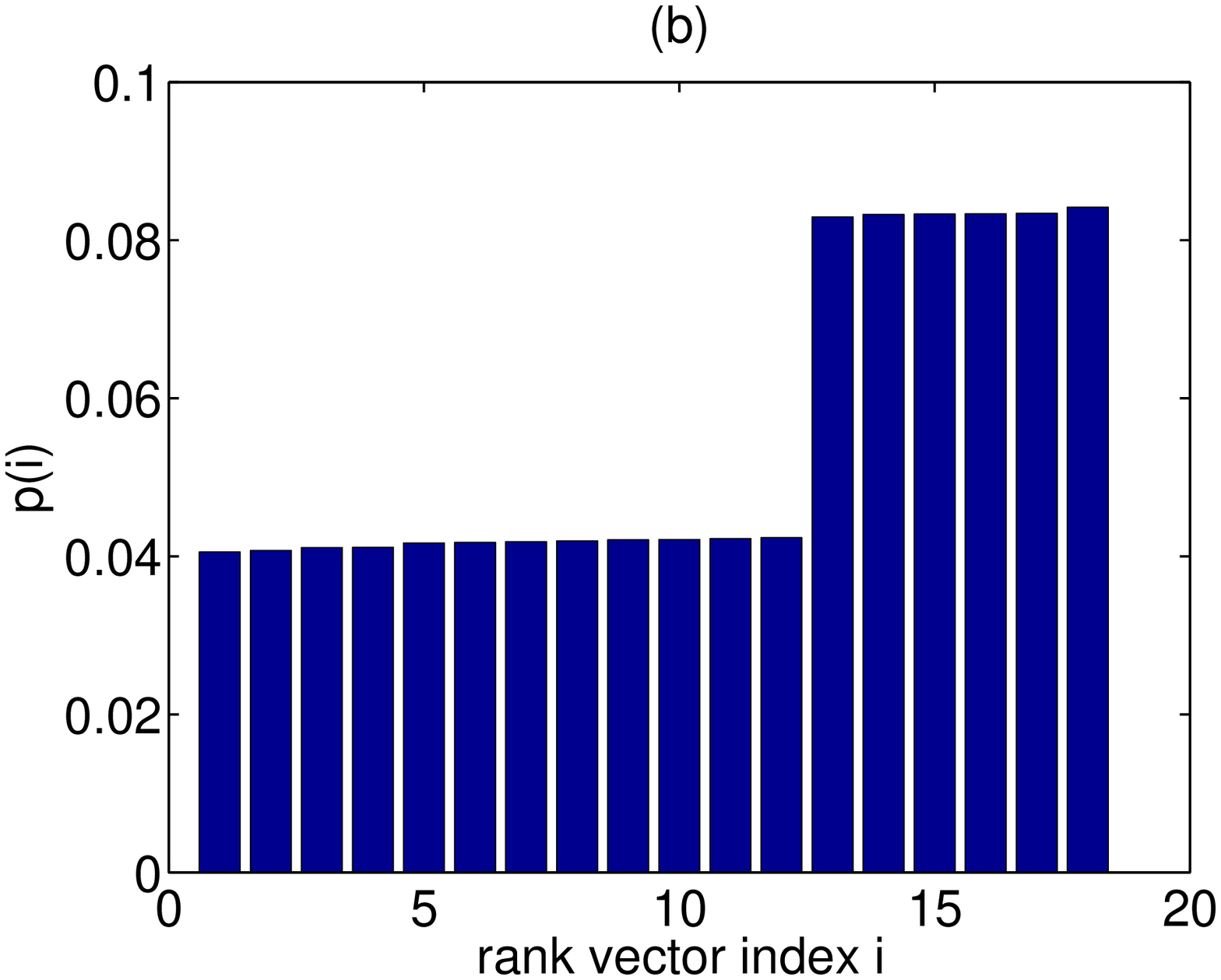}}
\caption{(a) Estimated pmf for the ranks of $[y_{t+1},\mathbf{y}_t]$ with $m_y=3$
(probabilities are in ascending order),
where the samples $y_t$ are from a uniform white noise time series of length $N=10^{16}$. (b) Same as in (a) but for the rank vector $[\hat{\mathbf{y}}_{t+1},\hat{\mathbf{y}}_t]$.}
 \label{fig:rankvecpmfuniform}
\end{figure}

The pmf of the rank vector derived from $[y_{t+1},\mathbf{y}_t]$ and the pmf of the rank vector $[\hat{\mathbf{y}}_{t+1},\hat{\mathbf{y}}_t]$ are shown in Fig.~\ref{fig:rankvecpmfuniform} for uniform white noise data and $m_y=3$. There are $(m_y+1)!=24$ equiprobable rank orders for $[y_{t+1},\mathbf{y}_t]$ (see Fig.~\ref{fig:rankvecpmfuniform}a) but only $m_y!\cdot\frac{m_y!}{(m_y-1)!}=18$ different vectors $[\hat{\mathbf{y}}_{t+1},\hat{\mathbf{y}}_t]$ are found, where $m_y!=6$ of them have about double probability, each corresponding to two distinct rank orders that could not be distinguished (Fig.~\ref{fig:rankvecpmfuniform}b).
This results in the underestimation of the Shannon entropy. Using $N=10^{16}$ samples and the ranks of $[y_{t+1},\mathbf{y}_t]$ we found $H=4.5846$ bits and using $[\hat{\mathbf{y}}_{t+1},\hat{\mathbf{y}}_t]$ we found $H=4.0865$ bits, while the true Shannon entropy is $H=-\log_2(1/24)=4.5850$.

Assuming a time step ahead $T>1$, there are two scenarios to follow for the future samples of system $Y$: a single sample at time $t+T$, $y_{t+T}$, or all the samples in the horizon of length $T$, which we denote as $\mathbf{y}_t^T=[y_{t+1},\ldots,y_{t+T}]$. In the first case the possible rank orders of $[y_{t+T},\mathbf{y}_t]$ are again $(m_y+1)!$ and for the second case the possible rank orders of $[\mathbf{y}_t^T,\mathbf{y}_t]$ are $(m_y+T)!$.
If instead we follow the form in STE and substitute $\hat{\mathbf{y}}_{t+T}$ to $\hat{\mathbf{y}}_{t+1}$, we have $m_y!\cdot\frac{m_y!}{(m_y-T)!}$ possible ranks for $[\hat{\mathbf{y}}_{t+T},\hat{\mathbf{y}}_t]$, which regards neither of the two joint vector forms. For example, for $m_y=3$ and $T=2$, using the form of STE $[\hat{\mathbf{y}}_{t+T},\hat{\mathbf{y}}_t]$ we have 36 possible rank orders, while
for the joint vector form with a single sample $T$ ahead, $[y_{t+T},\mathbf{y}_t]$,
the possible rank orders are 24 and for all $T$ samples ahead, $[\mathbf{y}_t^T,\mathbf{y}_t]$, they
are 120. For uniform white noise data, the true entropy for the ($m+1$)-dimensional augmented vector
is $H=4.5850$ and for ($m+T$)-dimensional augmented vector is $H=6.9069$. Using the same estimation setup as before, we estimated correctly these entropies as $H=4.5847$ and $H=6.9056$ using the rank orders of $[y_{t+T},\mathbf{y}_t]$ and $[\mathbf{y}_t^T,\mathbf{y}_t]$, respectively. Using the rank vector $[\hat{\mathbf{y}}_{t+T},\hat{\mathbf{y}}_t]$
we estimate $H=4.9709$, which constitutes overestimation if considering a single future sample at $T$ time
steps ahead and underestimation if considering all the samples in the $T$ time steps ahead.

For the future response at time $T$, we use the vector $\mathbf{y}_t^T$ of all the
samples in the future horizon $t+1,\ldots,t+T$. Thus we propose to substitute $\hat{\mathbf{y}}_{t+T}$
in STE of eq.(\ref{eq:STEH}) by $\hat{\mathbf{y}}^T_{t}=[\hat{y}_{t+1},\ldots,\hat{y}_{t+T}]$, the ranks of
$\mathbf{y}_t^T=[y_{t+1},\ldots,y_{t+T}]$ in the augmented vector $[\mathbf{y}_t^T,\mathbf{y}_t]$.
The proposed measure of transfer entropy on rank vectors (TERV) for $T$ steps ahead is
%----------------------------------------------------------------------
\begin{eqnarray}
\label{eq:TERVH}
& & \mbox{TERV}^T_{X\rightarrow Y} = \\
& & -H(\hat{\mathbf{y}}^T_{t},\hat{\mathbf{x}}_t,\hat{\mathbf{y}}_t) + H(\hat{\mathbf{x}}_t,\hat{\mathbf{y}}_t) + H(\hat{\mathbf{y}}^T_{t},\hat{\mathbf{y}}_t) -H(\hat{\mathbf{y}}_t). \nonumber
\end{eqnarray}
%----------------------------------------------------------------------
Since $\hat{\mathbf{y}}^T_{t}$ appears together with $\hat{\mathbf{y}}_{t}$ in
the two entropy terms in eq.(\ref{eq:TERVH}), one can define
$[\hat{\mathbf{y}}^T_{t}, \hat{\mathbf{y}}_{t}]$ as the ranks of the augmented vector $[y_{t+T},y_{t+T-1},\ldots,y_{t+1},\mathbf{y}_t]$. This is actually equivalent of
taking the ranks of $\hat{\mathbf{y}}_{t}$ independently (as they appear in the second
and forth term in eq.(\ref{eq:TERVH})).

The use of all the ranks for times $t+1,\ldots,t+T$ aims at capturing
the effect of $X$ on the evolution of the time series of $Y$ up to $T$ time steps ahead.
Similar reasoning for $T>1$ was used for other information flow measures \cite{Palus01,Vejmelka08}
and we have used $T>1$ also for TE in \cite{Papana08b}.
The TERV measure is the direct analogue to TE using ranks and extends the measure of
information flow from $X$ to $Y$ at time $t$ for a range of $T$ time steps ahead $t$.
The use of $\mathbf{y}_t^T$ instead of $y_{t+T}$ increases the dimension of the joint space from
$m_x+m_y+1$ to $m_x+m_y+T$ and can affect the stability of the estimation. However, the results of
a simulation study were in favor of $\mathbf{y}_t^T$ against $y_{t+T}$ for both TE and TERV.

Finally, we note that when a lag $\tau_y>1$ is used for the state space reconstruction of $y_t$, there are up to $m_y!\cdot m_y!$ different rank vectors $[\hat{\mathbf{y}}_{t+T},\hat{\mathbf{y}}_t]$ in the computation of STE. On the other hand, for TERV there are $(T+m_y)!$ different rank vectors $[\hat{\mathbf{y}}^T_{t},\hat{\mathbf{y}}_t]$. Thus for $\tau_y>1$, the distortion of the domain of the rank vectors by STE may be large, e.g. for $\tau_y=2$ and $T=1$, the pmfs and entropies are computed on $(m_y+1)!$ different rank orders for TERV and $m_y!\cdot m_y!$ for STE.

%%%%%%%%%%%%%%%%%%%%%%%%%%%%%%%%%%%%%%%%%%%%%%%%%%%%%%%%%%%%%%%%%
\section{Estimation of information measures on simulated systems}
\label{sec:simulations}
%%%%%%%%%%%%%%%%%%%%%%%%%%%%%%%%%%%%%%%%%%%%%%%%%%%%%%%%%%%%%%%%%

As it was shown for the example of uniform white noise the distortion of the domain of the rank vectors $[\hat{\mathbf{y}}^T_{t},\hat{\mathbf{y}}_t]$ using the rank vectors $[\hat{\mathbf{y}}_{t+T},\hat{\mathbf{y}}_t]$ instead has a direct effect on the estimation of entropy. While for uncoupled systems $X$ and $Y$ the entropy terms involving $[\hat{\mathbf{y}}_{t+T},\hat{\mathbf{y}}_t]$ cancel out in the expression of TERV (and respectively for STE), in the presence of coupling some bias is introduced in the estimation of the coupling measure by STE. Using TERV instead this bias is removed.

We compare the estimation of coupling (strength and direction) with the measures TE, STE and TERV on simulated systems.
We first standardize each time series to have mean zero and standard deviation one, and this allows us to define a fixed radius for all systems in the computation of TE, which we set $r=0.15$. This choice is a trade-off of having enough points within a distance $r$ to assure stable estimation of the point distribution and maintaining small neighborhoods to preserve details of the point distribution. Still, for high-dimensional points, even this radius may be insufficient to provide
stable estimation.

\begin{figure}[htb!]
\centering
\hbox{\includegraphics[height=3.5cm]{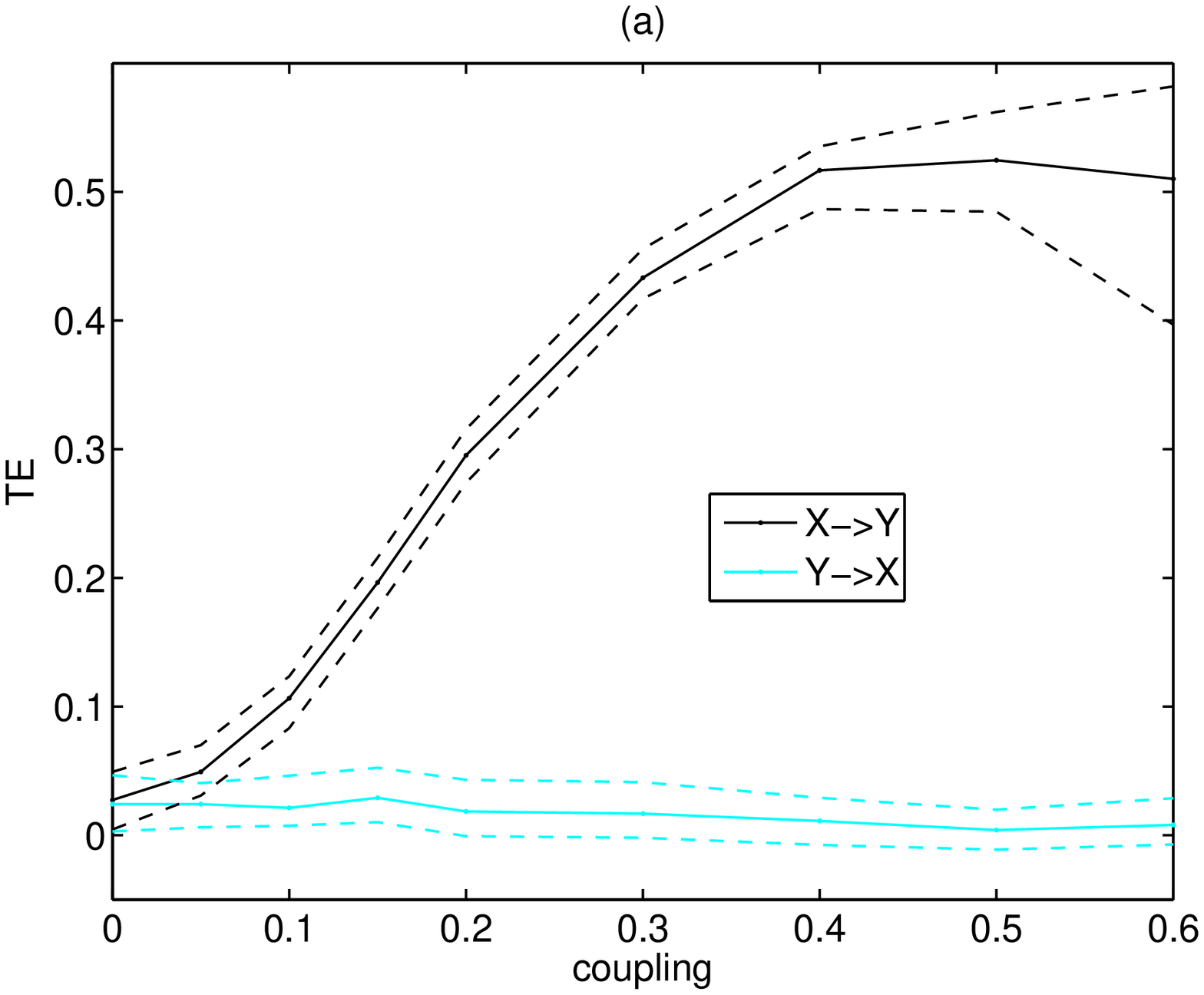}
\includegraphics[height=3.5cm]{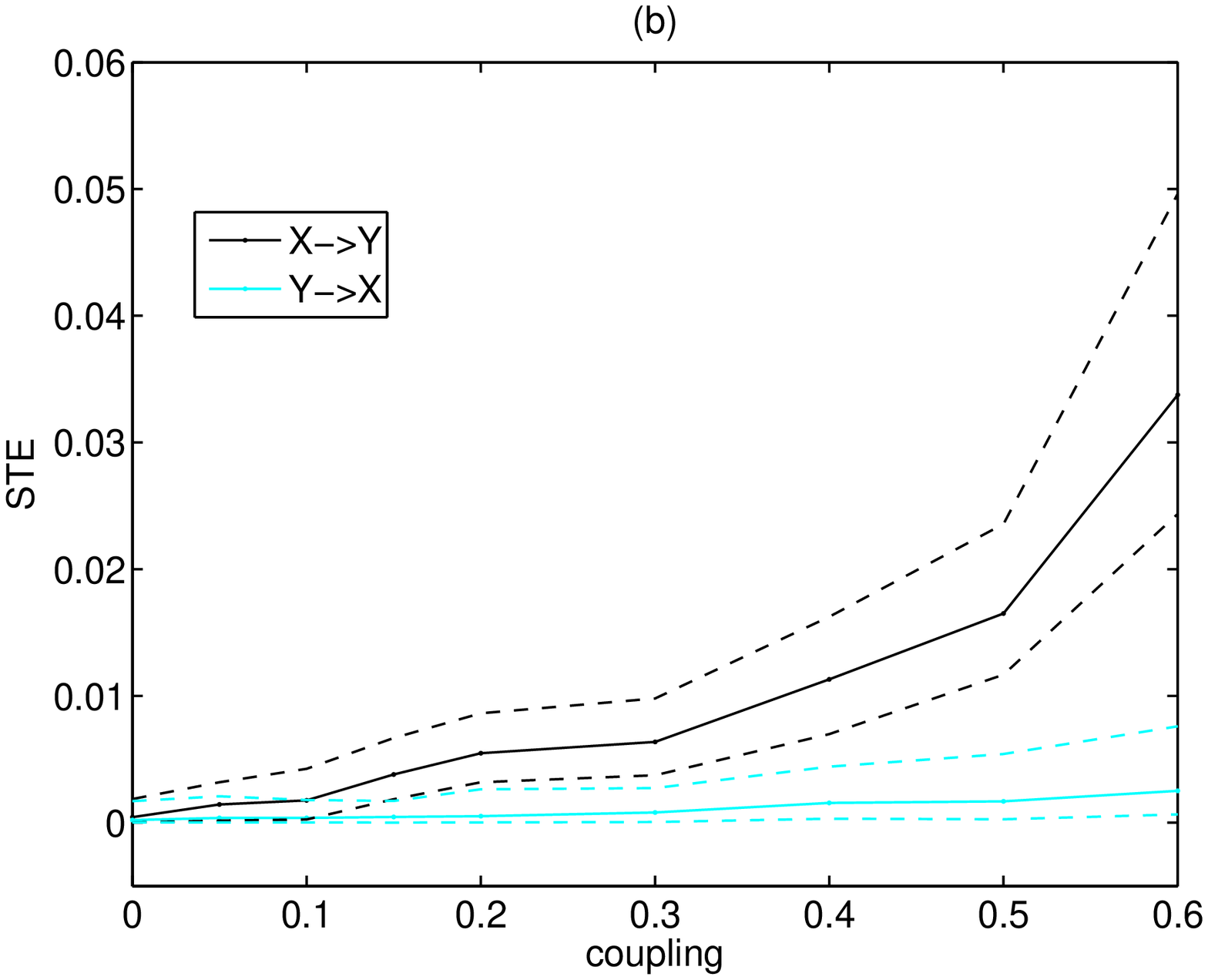}}
\hbox{\includegraphics[height=3.5cm]{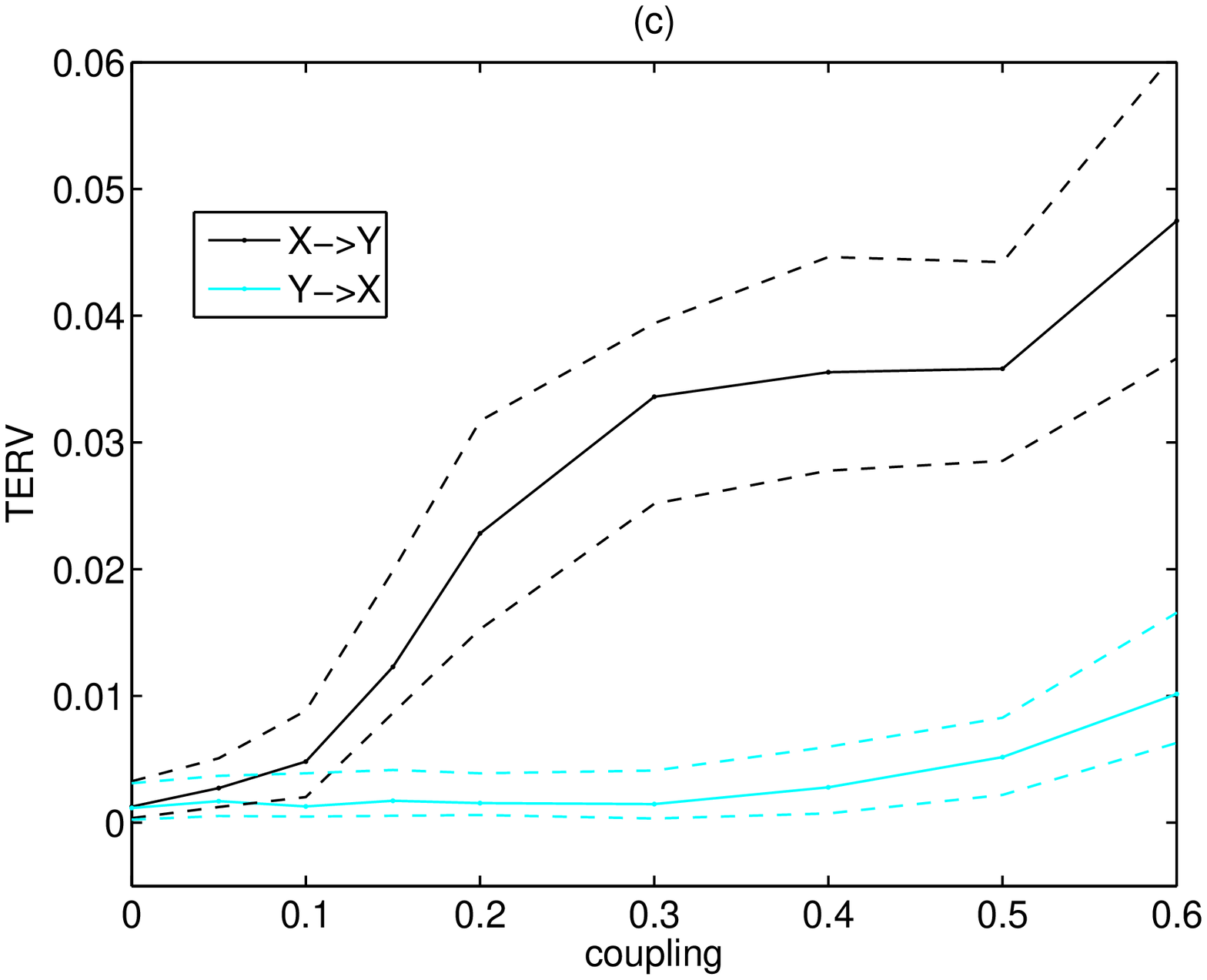}
\includegraphics[height=3.5cm]{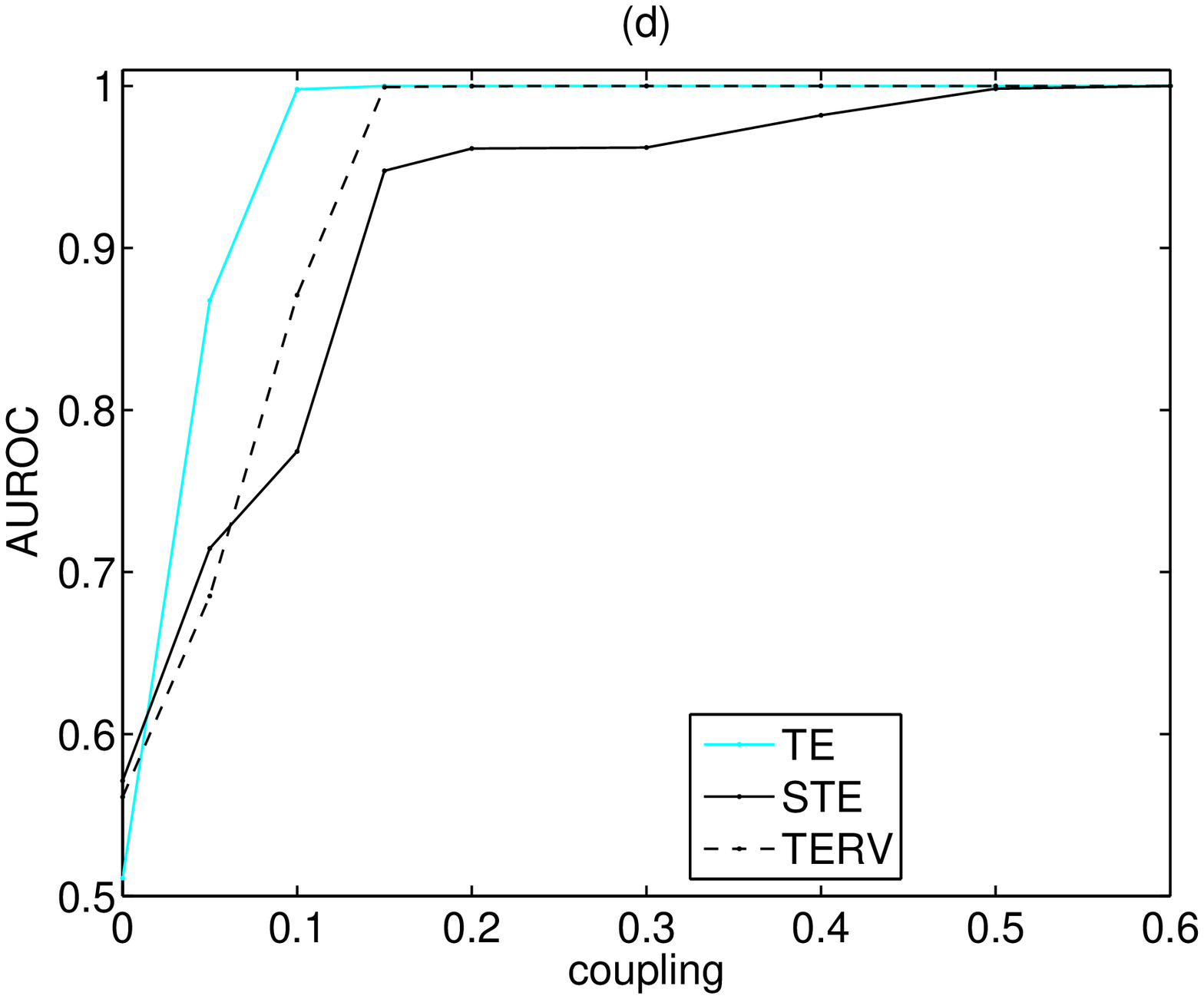}}
\caption{(a) Median (solid line) and $12.5\%$ and $87.5\%$ percentiles (dashed lines) of TE computed on 100 noise-free realizations of length $N=1024$ from the system of two unidirectionally coupled Henon maps for varying coupling strengths. The other parameters are $T=1$, $\tau_x=\tau_y=1$ and $m_x=m_y=2$. The direction $X\rightarrow Y$ is shown with black lines and $Y\rightarrow X$ with grey (online cyan) lines, as shown in the legend. (b) Same as (a) but for STE. (c) Same as (a) but for TERV. (d) AUROC computed on the 100 realizations for each of the two directions and for the measures TE, STE and TERV, as given in the legend.}
 \label{fig:unidHennl0n1024m12m22}
\end{figure}

We start with two unidirectionally coupled Henon maps \cite{Schiff96}
\begin{eqnarray*}
x_{t+1} & = & 1.4  - x^2_t + 0.3x_{t-1} \\
y_{t+1} & = & 1.4 - cx_t y_t + (1-c)y^2_t + 0.3y_{t-1}
\end{eqnarray*}
with coupling strengths $c=$0,0.05,0.1,0.15,0.2,0.3,0.4,0.5 and 0.6.
The results on the coupling measures TE, STE and TERV for $T=1$, $\tau_x=\tau_y=1$ and $m_x=m_y=2$ are shown for 100 noise-free bivariate time series of length $N=1024$ in Fig.~\ref{fig:unidHennl0n1024m12m22}.
TE has the smallest variance and it seems to give the best detection of the correct direction of coupling even for very weak coupling, whereas STE performs worst. To quantify the level of discrimination of the correct direction of information flow, $X\rightarrow Y$, often the net information flow is used, defined as the difference of the coupling measure in the two directions. Here, we assess the level of discrimination in a statistical setting by computing the area under the receiver operating characteristic (ROC) curve on the 100 coupling measure values for each direction, which we denote AUROC (e.g. see \cite{Hand01}). For uncoupled systems, we expect that AUROC be close to 0.5. For an information flow measure to detect coupling with great confidence AUROC has to be close to 1. In Fig.~\ref{fig:unidHennl0n1024m12m22}d, the AUROC shows that TE detects coupling with great confidence and obtains AUROC=1 for as low coupling strength as $c=0.1$, followed by TERV reaching the same level of confidence at $c=0.15$, while STE reaches this level only at strong coupling ($c=0.5$).

\begin{figure}[htb!]
\centering
\hbox{\includegraphics[height=3.5cm]{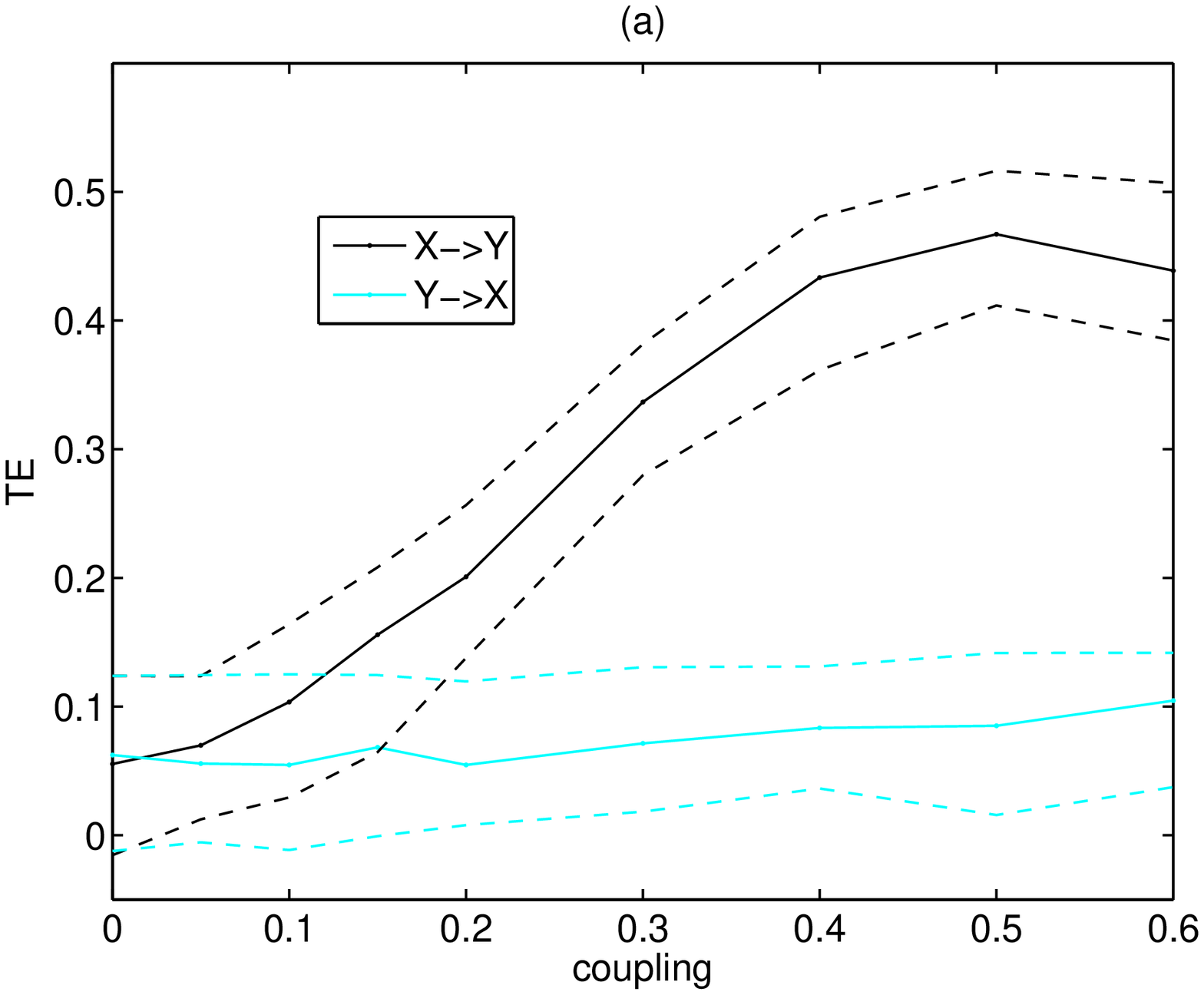}
\includegraphics[height=3.5cm]{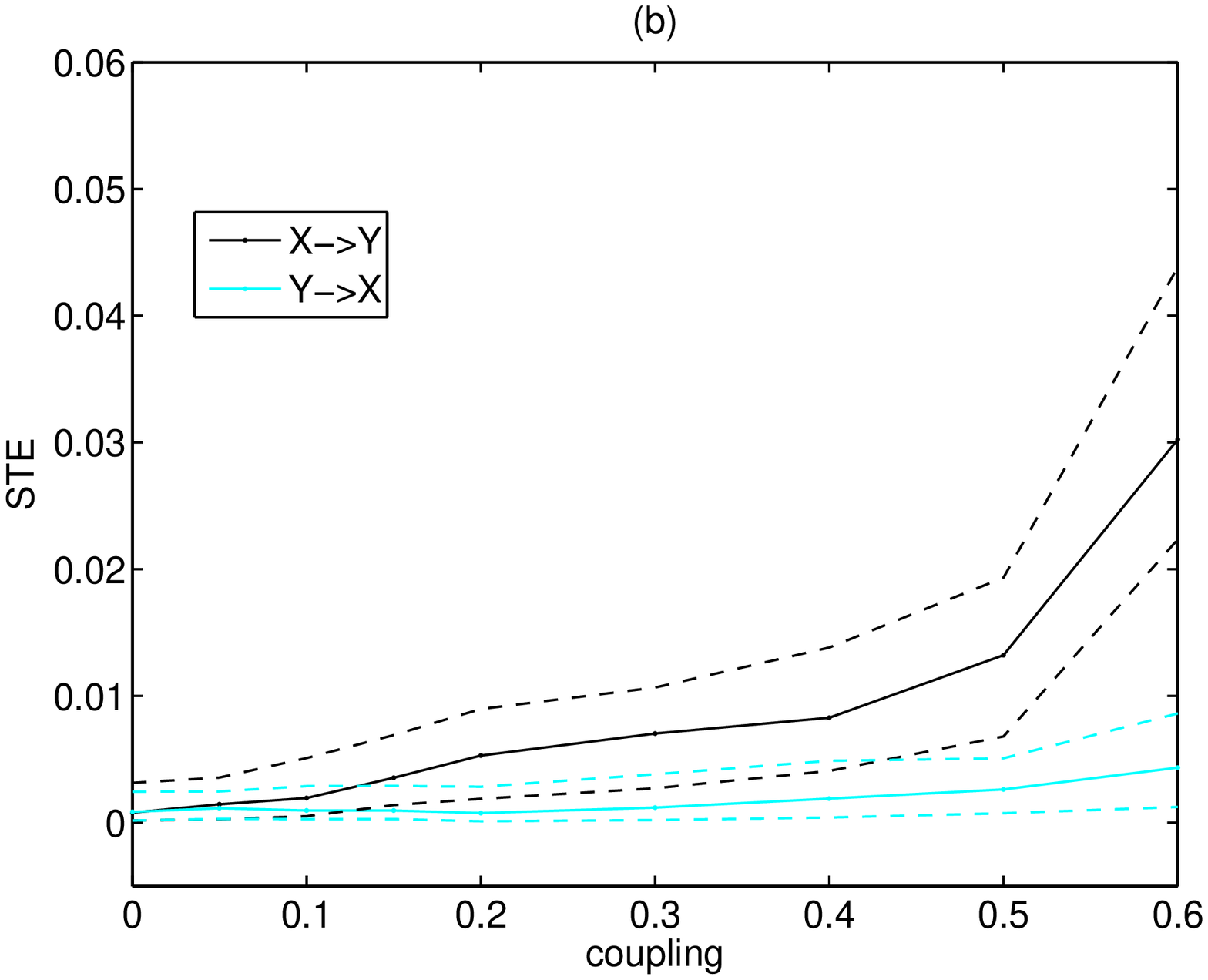}}
\hbox{\includegraphics[height=3.5cm]{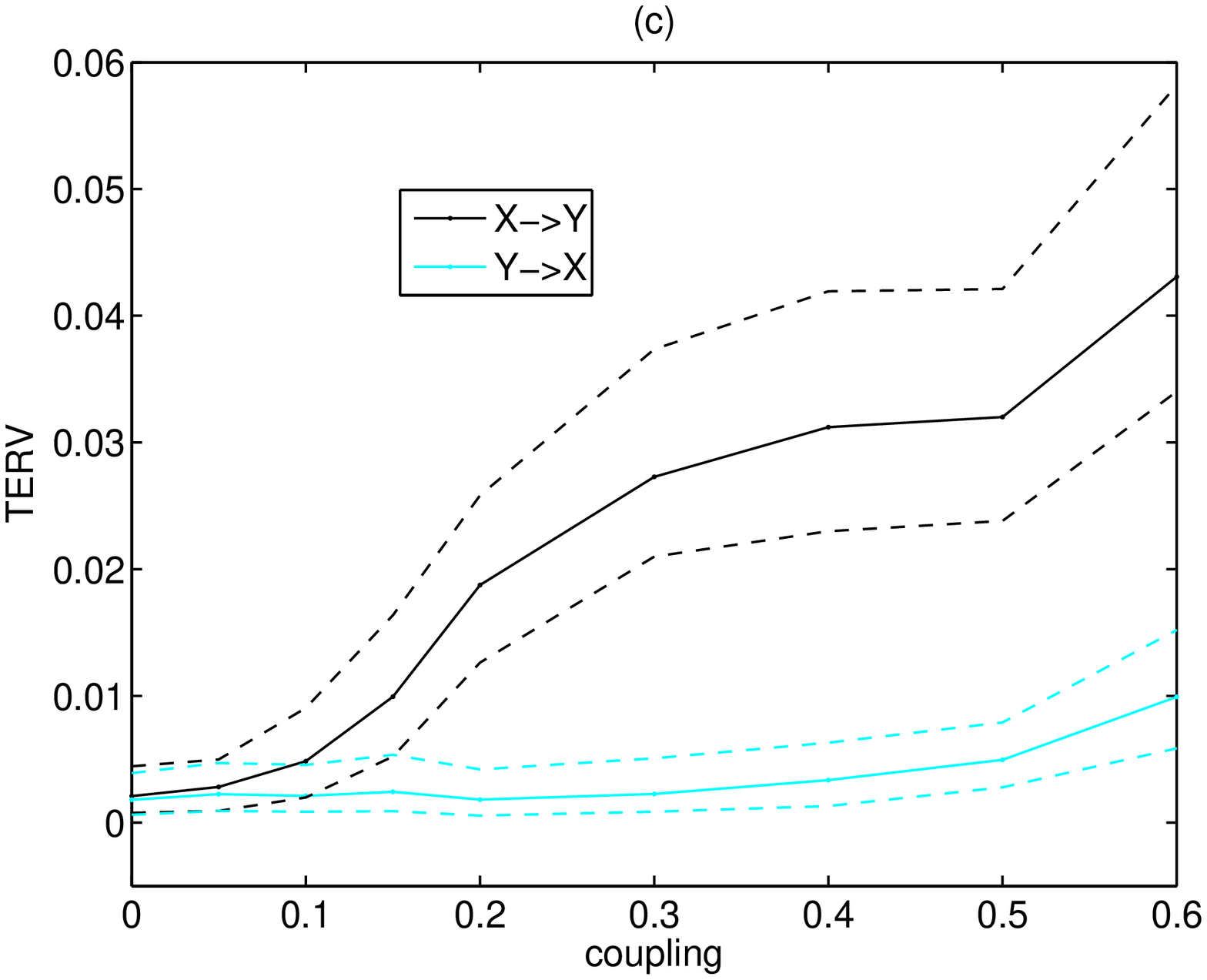}
\includegraphics[height=3.5cm]{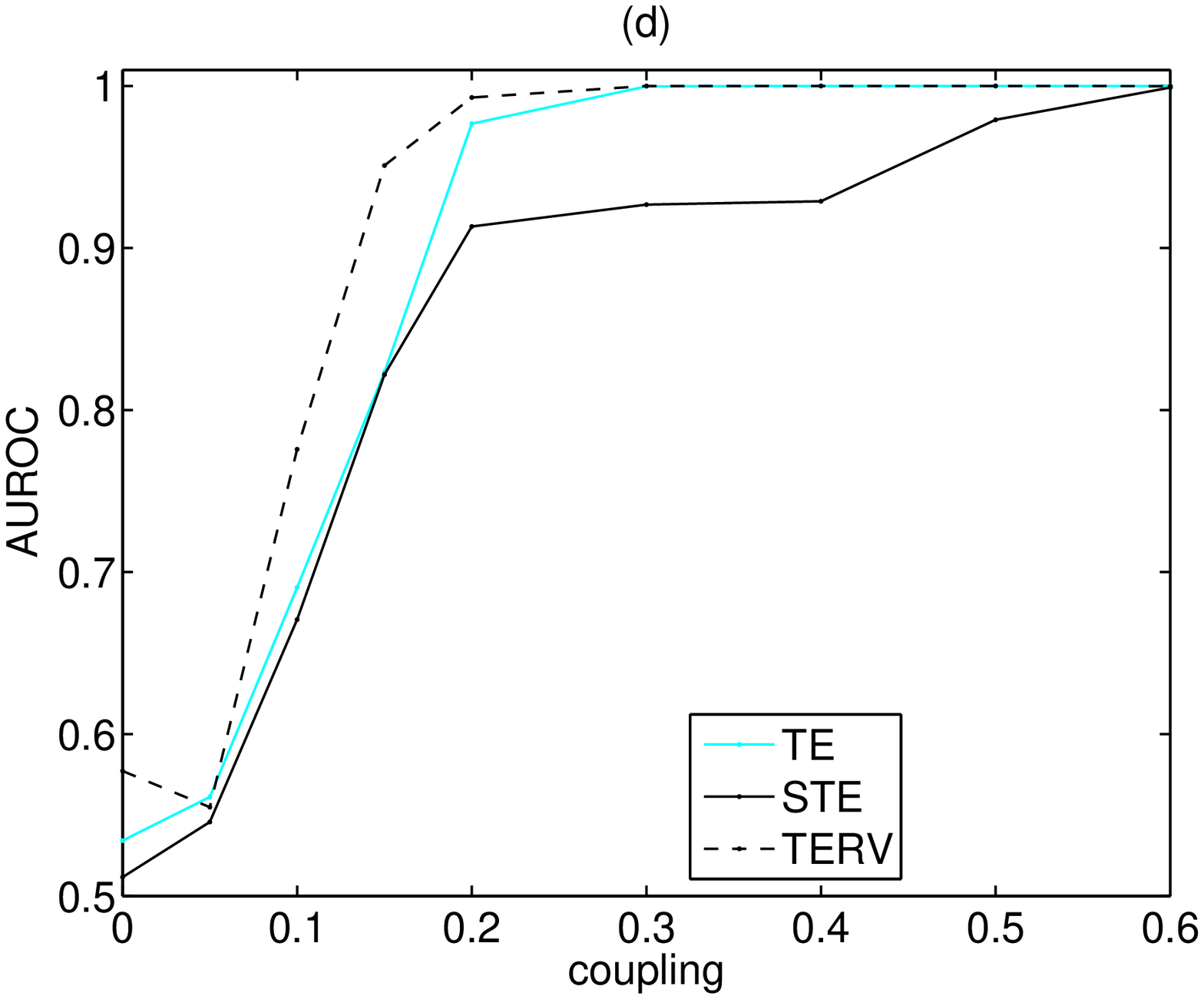}}
\caption{As Fig.~\ref{fig:unidHennl0n1024m12m22}, but with Gaussian white noise with standard deviation 0.2 added to the standardized time series.}
 \label{fig:unidHennl20n1024m12m22}
\end{figure}

The performance of the coupling measures changes in the presence of noise. For the same setup as that in Fig.~\ref{fig:unidHennl0n1024m12m22}, but adding to the bivariate time series 20\% Gaussian white noise, we observe that TERV performs best, followed by TE and having STE with the smallest increase in the direction $X \rightarrow Y$ with the coupling strength (see Fig.~\ref{fig:unidHennl20n1024m12m22}).
TERV has smaller variance than TE for small coupling strengths, which increases its discriminating power. Comparing with the noise-free case in Fig.~\ref{fig:unidHennl0n1024m12m22}, TERV does not seem to be much affected by the addition of noise, but TE does not have the same robustness to noise. It is noted that using $r=0.15$ was particularly suitable to maintain some stability in TE on the noisy data. The same simulations for $r=0.1$ gave much worse results \cite{Kugiumtzis09}. The AUROC curves in Fig.~\ref{fig:unidHennl0n1024m12m22}d are ordered with TERV giving the highest and STE the lowest AUROC for all coupling strengths.

We have estimated TE, STE and TERV on the coupled Henon system for different settings of embedding dimensions,
time steps ahead $T$, time series length $N$ and noise level. For $N$ small and $m_y$ large and mostly for noisy time series, the computation of TE was unstable due to the lack of points within the given radius. This explains that TE
has smaller variance for larger time series, and consequently better discrimination in the two directions of coupling. The measures do not vary much with the embedding dimensions and STE shows the largest dependency, especially when $m_x > m_y$. The best results for all measures were obtained for $m_x=m_y$.

\begin{figure}[htb!]
\centering
\hbox{\includegraphics[height=3.5cm]{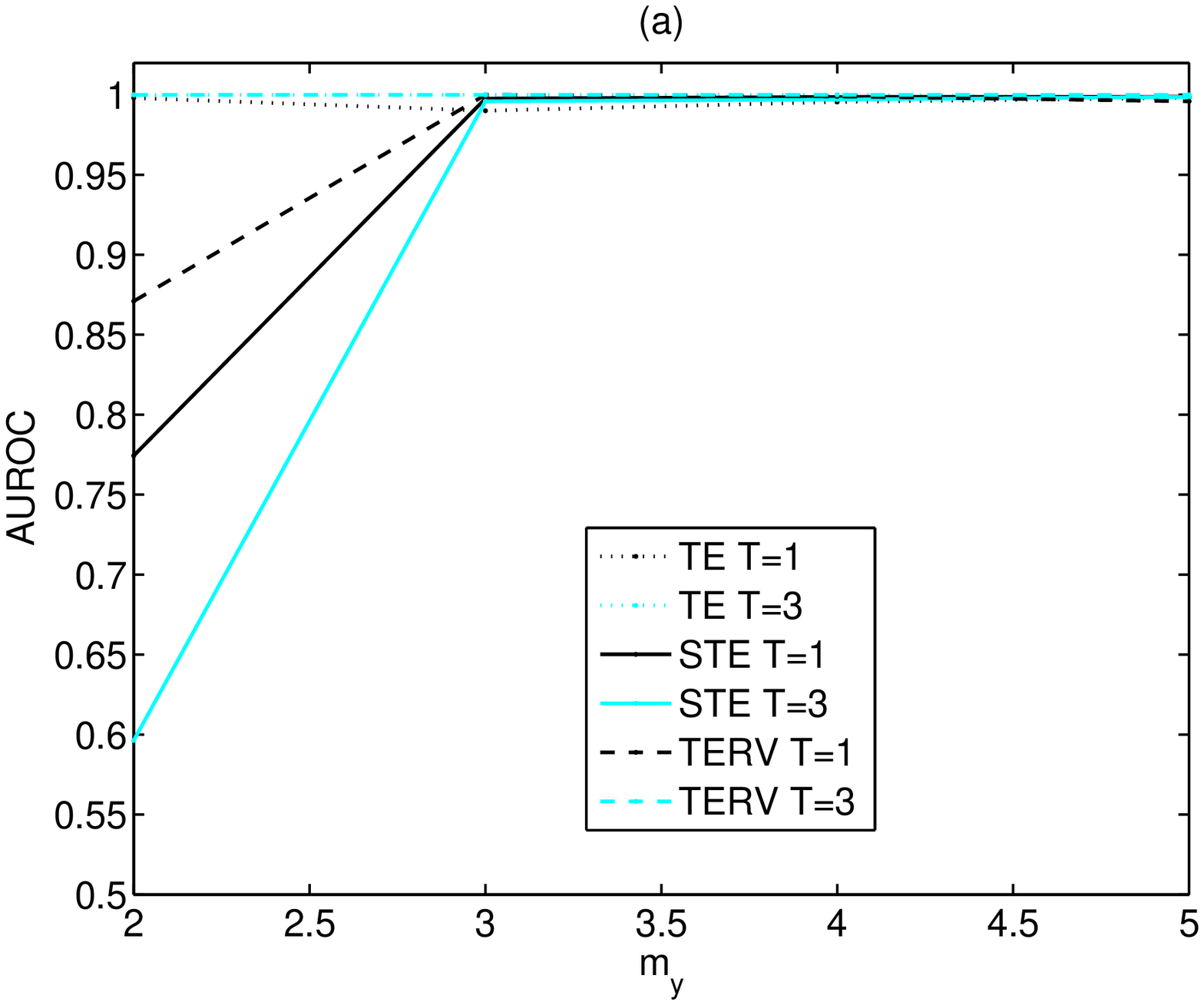}
\includegraphics[height=3.5cm]{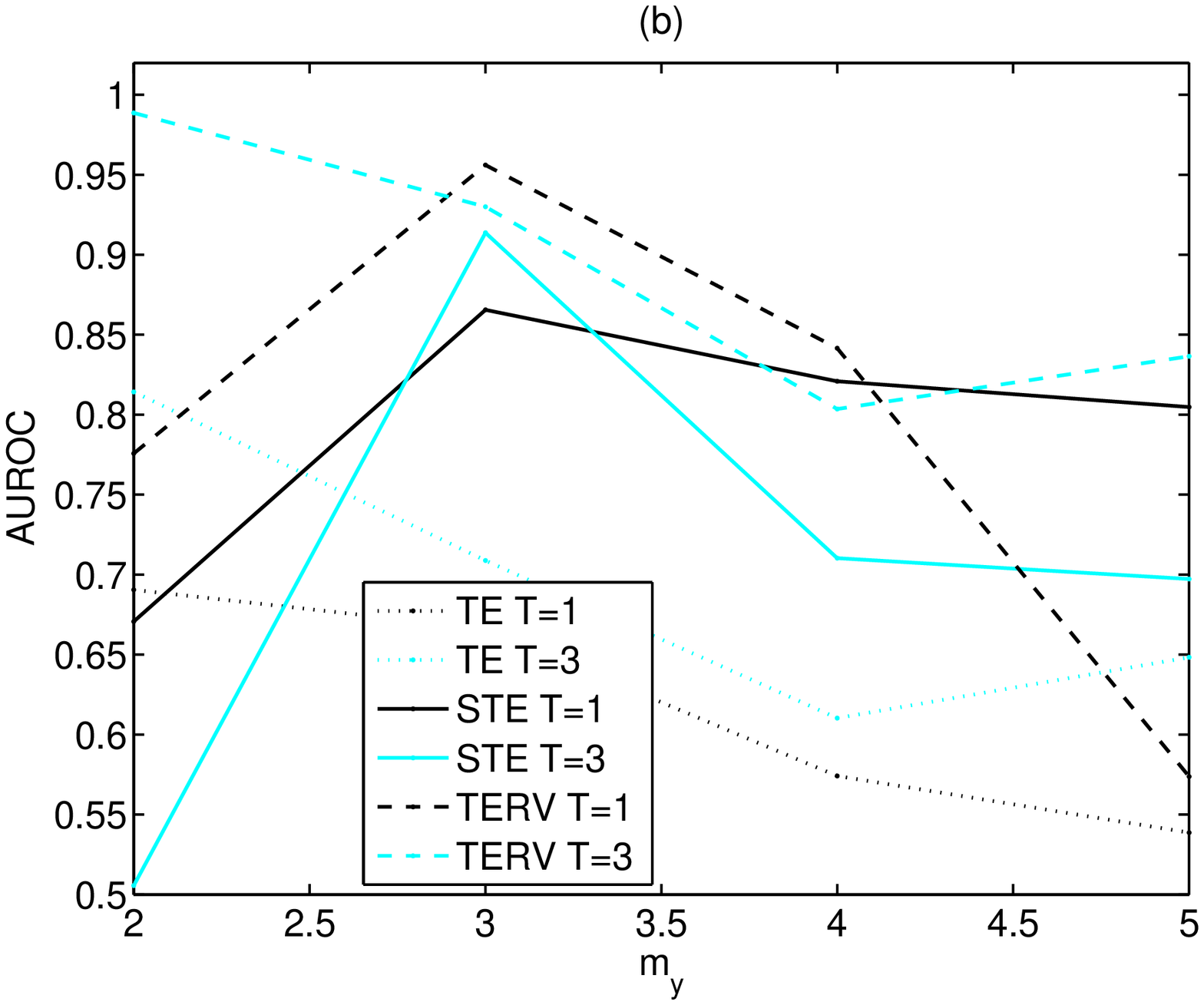}}
\hbox{\includegraphics[height=3.5cm]{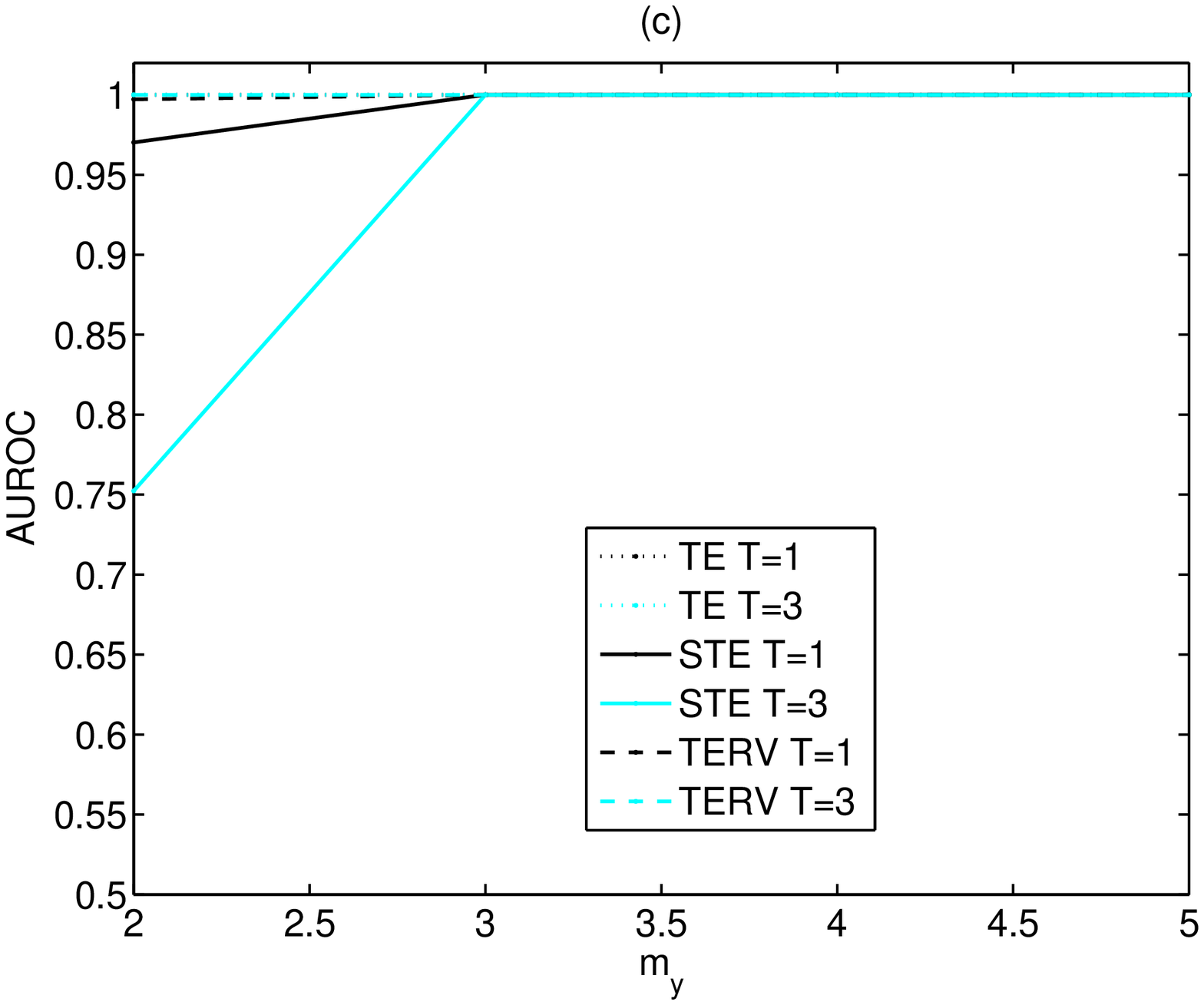}
\includegraphics[height=3.5cm]{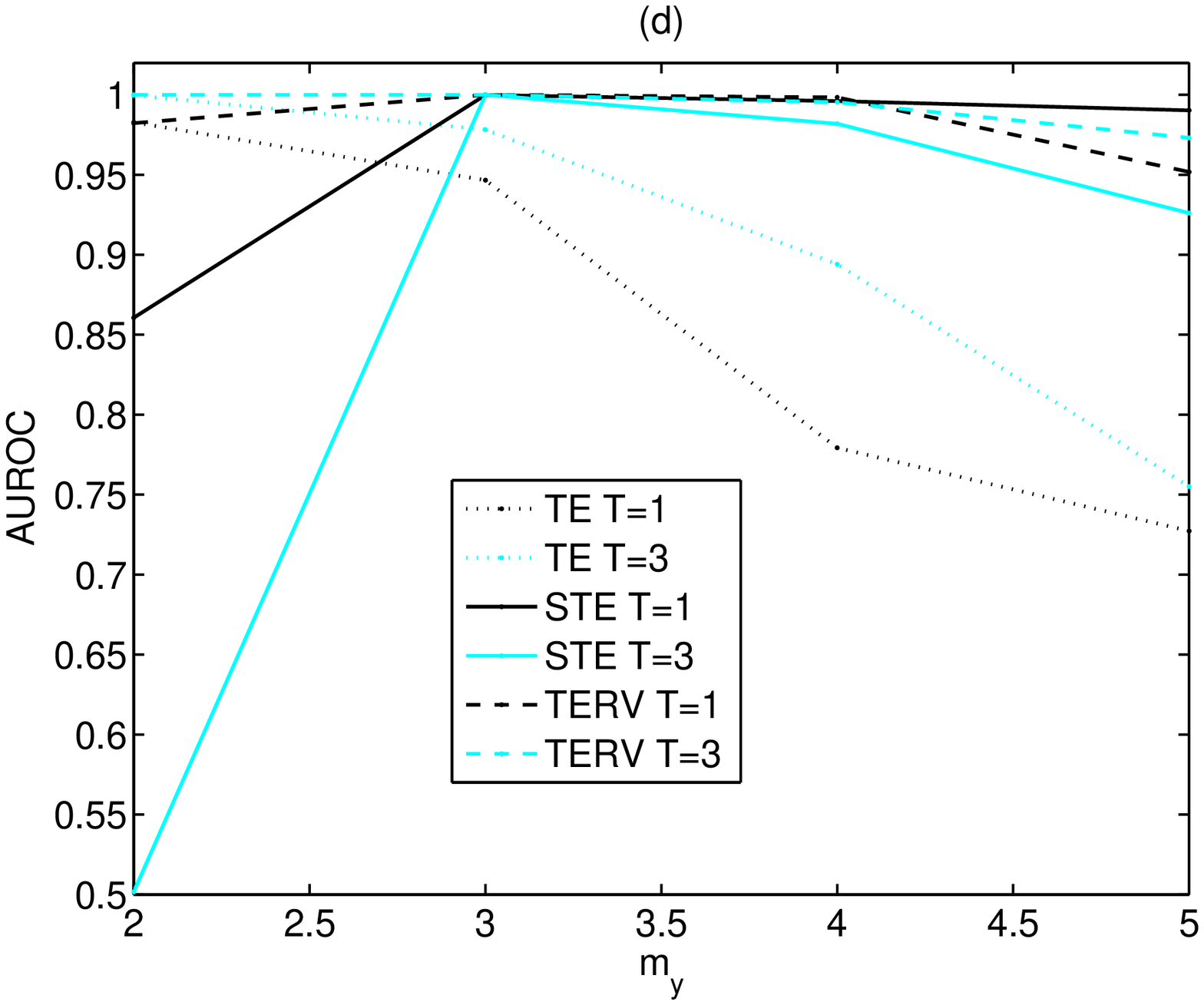}}
\caption{(a) AUROC computed for different $m_y$ ($m_x=m_y$) on 100 realizations of the weakly coupled Henon system ($c=0.1$) for each of the two directions, for the measures TE, STE and TERV and for time steps ahead $T$, as given in the legend. The time series are noise-free and $N=1024$. (b) As in (a) but for 20\% additive Gaussian white noise. (c) and (d) are as in (a) and (b) but for $N=4096$.}
 \label{fig:unidHendiversesetup}
\end{figure}

In Fig.~\ref{fig:unidHendiversesetup}, the AUROC is shown for the three measures as a function of $m_y$, where $m_x=m_y$, for very weak coupling ($c=0.1$), and for one and three steps ahead, small and large time series and for noise-free and noisy Henon data.
It seems that estimating the information flow for $T=3$ increases the detection of correct direction of weak coupling for TE and TERV, but not for STE. For the noise-free data, the differences in AUROC among the three measures are small and all measures reach the highest level of discrimination of the two directions when $m_y>2$. For $m_y=2$, AUROC=1 is still reached by TE and TERV but only with $T=3$, whereas the AUROC is much smaller for STE regardless of $T$. This pattern is the same for both small and large $N$. For noisy data, all measures perform worse and their AUROC shows
strong dependence on both $m_y$ and $T$. Similarly to the noise-free case, the AUROC is larger for $T=3$ than for $T=1$ for TE and TERV, but not for STE. The AUROC of TE decreases with $m_y$ and for $m_y>2$ is lower than the AUROC for both STE and TERV. TERV obtains the highest AUROC values with $T=3$ giving the overall best results. The above results are consistent for the two time series lengths shown in Fig.~\ref{fig:unidHendiversesetup}b and d with AUROC values increasing from $N=1024$ to $N=4096$.

\begin{figure}[htb!]
\centering
\hbox{\includegraphics[height=3.5cm]{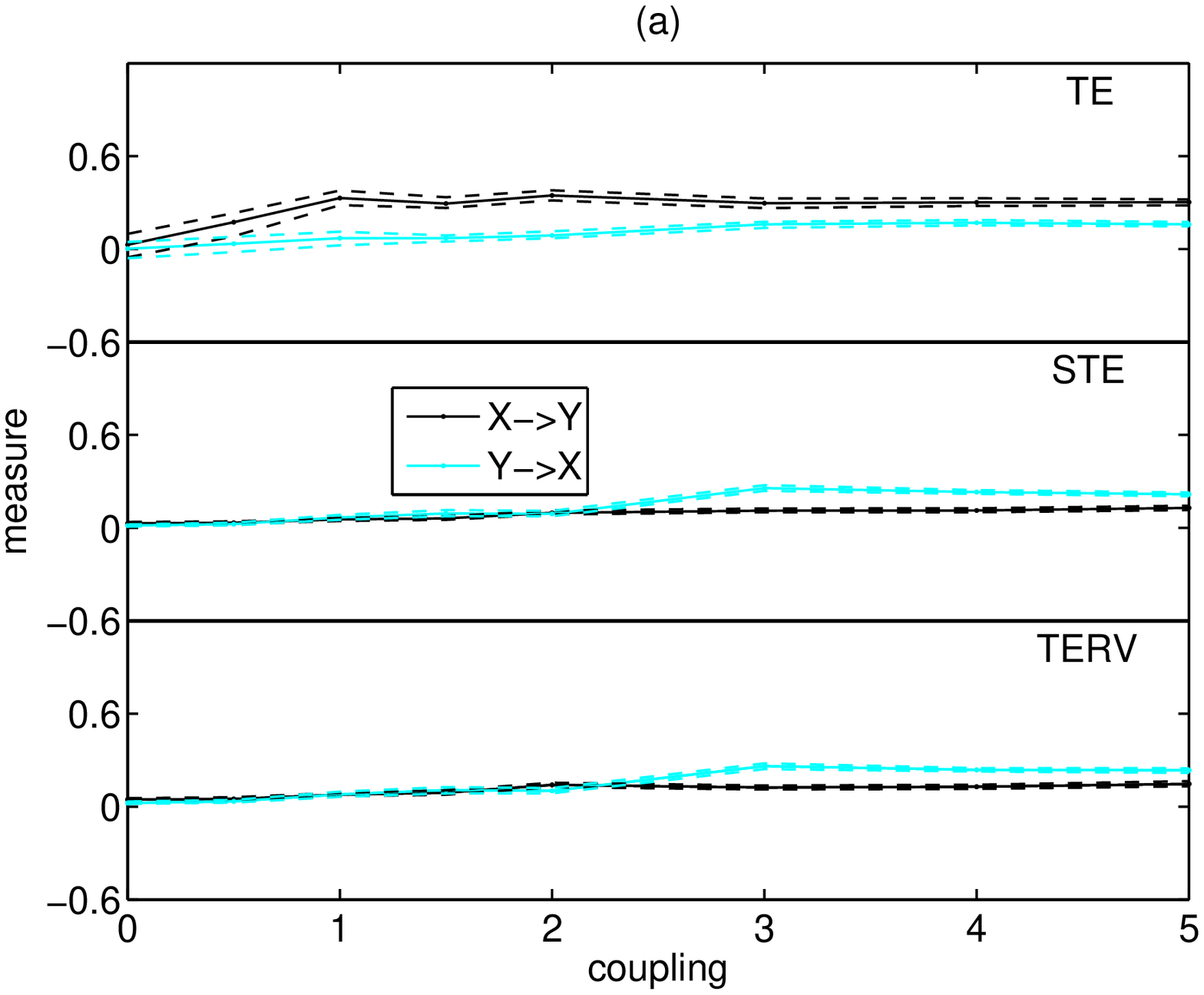}
\includegraphics[height=3.5cm]{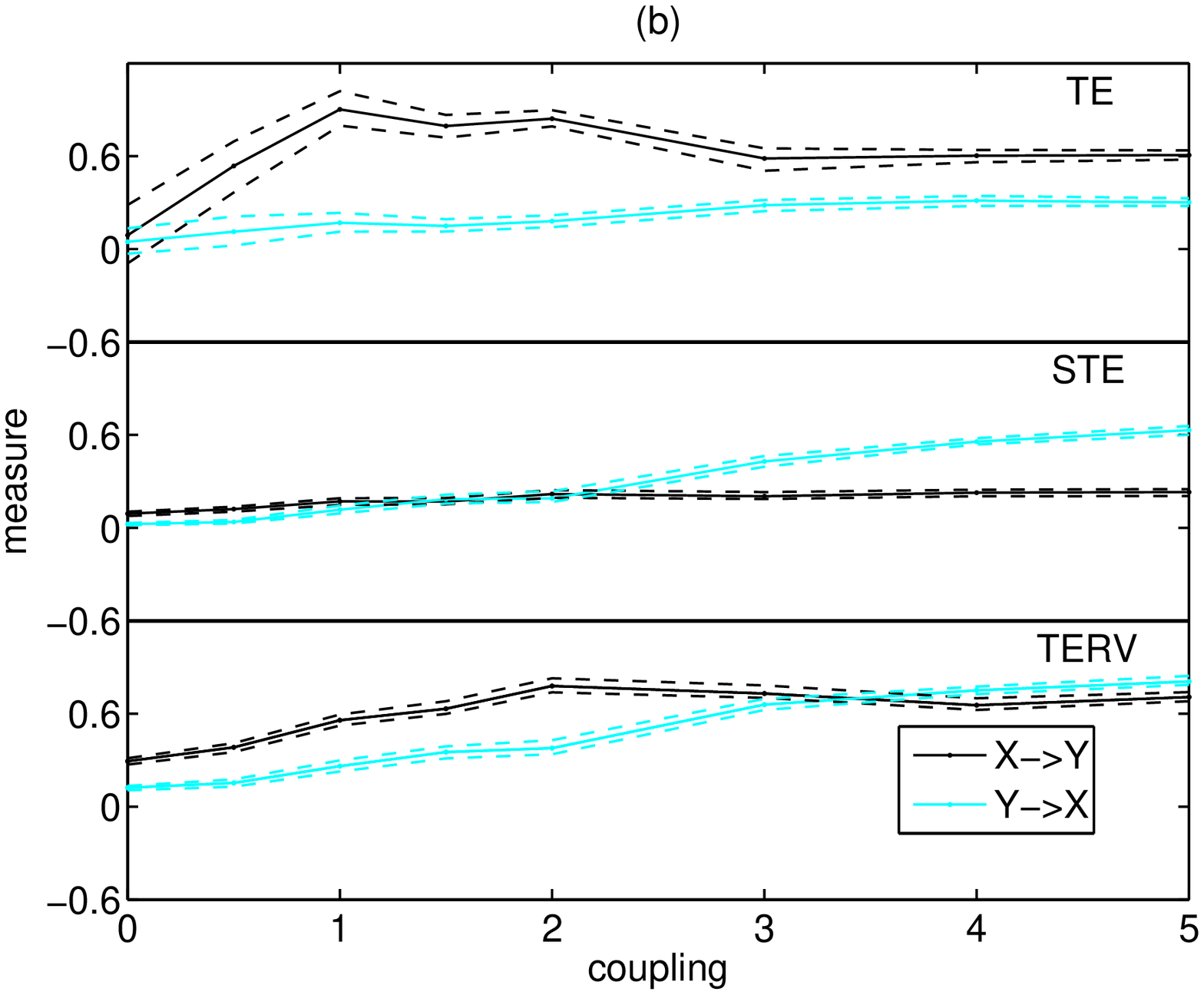}}
\hbox{\includegraphics[height=3.5cm]{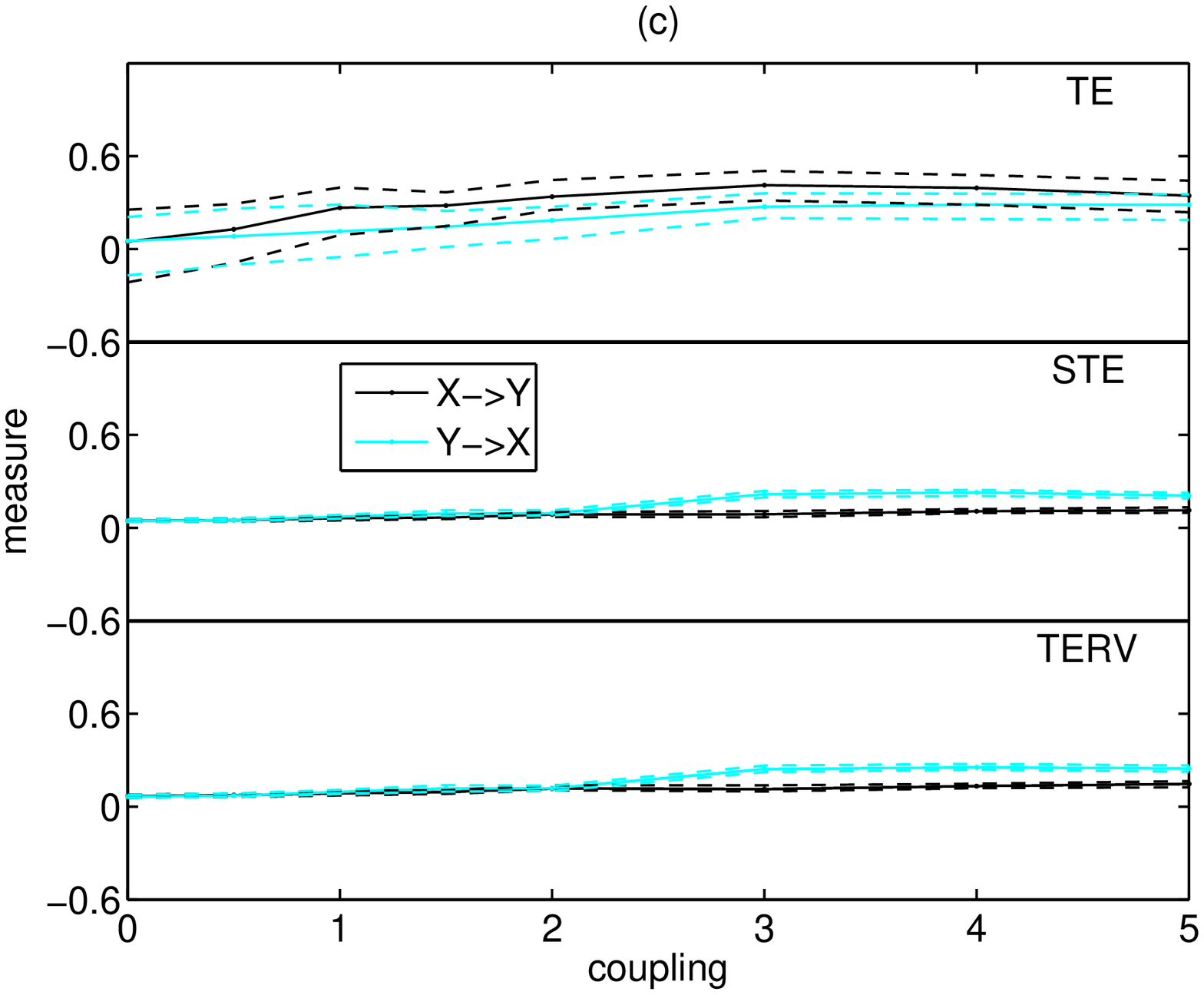}
\includegraphics[height=3.5cm]{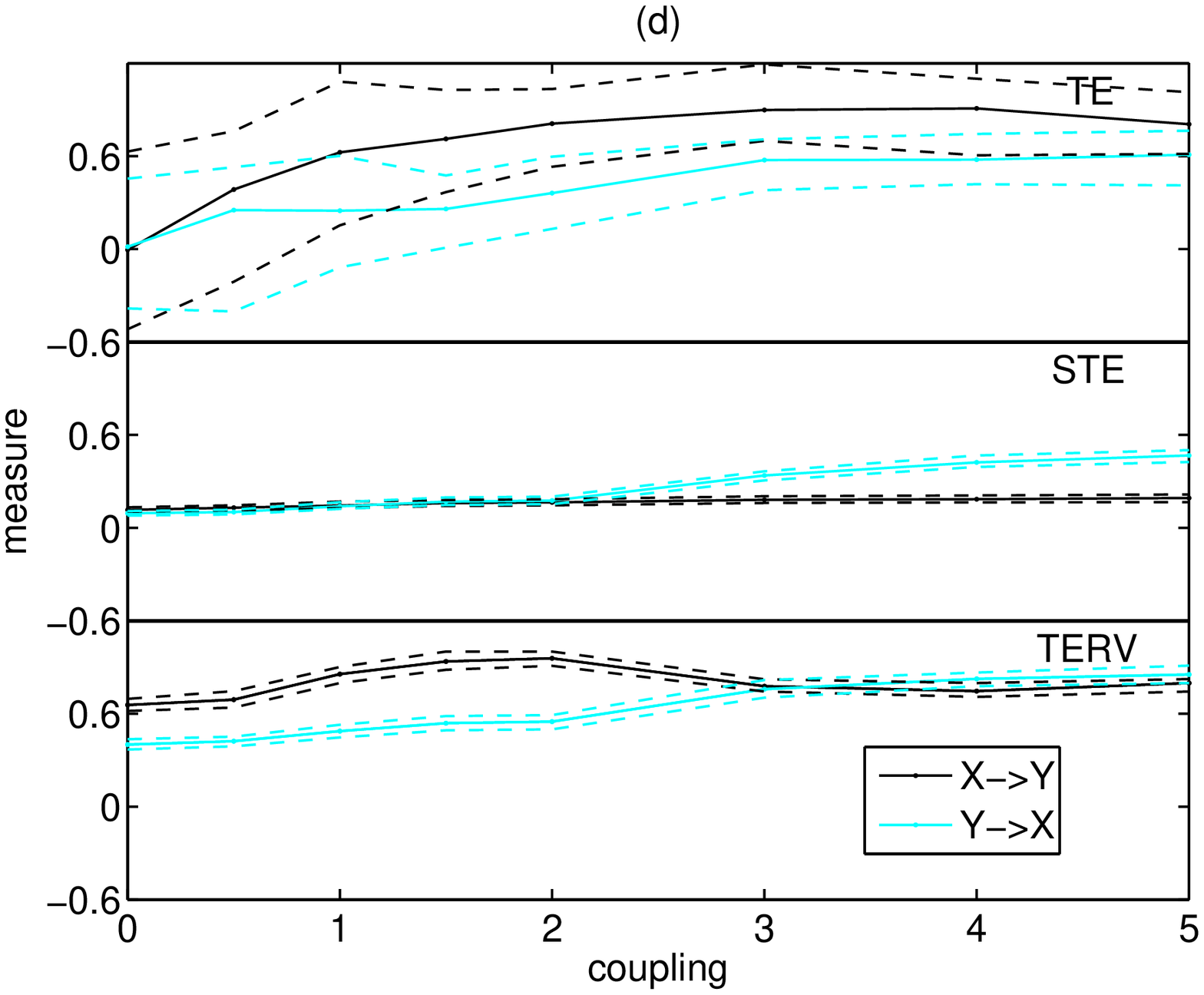}}
\caption{(a) Median (solid line) and $12.5\%$ and $87.5\%$ percentiles (dashed lines) of each of the three coupling measures computed on 100 noise-free realizations of length $N=1024$ from the R{\"o}ssler--Lorenz system for varying coupling strengths. The other parameters are $T=1$, $\tau_x=\tau_y=1$ and $m_x=m_y=3$. The direction $X\rightarrow Y$ is shown with black lines and $Y\rightarrow X$ with grey (online cyan) lines, as shown in the legend, and the TE measure is at the top panel, STE in the middel, and TERV at the low panel. (b) As in (a) but for $T=3$. (c) and (d) are as in (a) and (b) but for 20\% additive Gaussian white noise.}
 \label{fig:unidRoeLor}
\end{figure}

Similar simulations have been run for a R{\"o}ssler system driving a Lorenz system given as (subscript 1 for R{\"o}ssler, 2 for Lorenz)
\[
  \begin{array}{ll}
    \dot{x}_1=-6(y_1+z_1)  &  \dot{x}_2=10(x_2+y_2)  \\
    \dot{y}_1=-6(x_1+0.2y_1) & \dot{y}_2=28x_2-y_2-x_2z_2+ cy_1^2 \\
    \dot{z}_1=-6(0.2+z_1(y_1-5.7)) & \dot{z}_2=x_2y_2-\frac{8}{3}z_2
  \end{array}
\]
for coupling strengths $c=0,0.5,1,1.5,2,3,4,5$ \cite{Quyen99}. The observed variables are $x_2$ and $y_2$ and the sampling time is $\tau_s=0.1\mbox{sec}$. Here, the results are more varying than for the Henon system. First, there is stronger dependence of all measures, and particularly STE and TERV, on the two embedding dimensions. Again the best results are obtained for $m_x=m_y$, while for $m_x<m_y$ the measure values in the correct direction $X \rightarrow Y$ are increased and for $m_x>m_y$ the opposite is observed leading to erroneous detection of direction of interaction. The effect of the different formation of the rank future vector in STE and TERV can be better seen for $T>1$ and for small embedding dimensions. As shown in Fig.~\ref{fig:unidRoeLor}, for $m_x=m_y=3$, while for $T=1$ both rank measures tend to give larger values in the opposite wrong direction $Y \rightarrow X$, for $T=3$ STE continues to give the same result but TERV points to the correct direction at least for intermediate values of $c$.

The rank measures suffer from positive bias that increases with $T$ and this is more obvious in TERV.  When the two systems have different complexity the bias tends to be larger in the direction from the less complex (R{\"o}ssler) to the more complex (Lorenz) system. For $T=3$ in Fig.~\ref{fig:unidRoeLor}b and d, the rank measure values are larger for the direction $X \rightarrow Y$ when $c=0$. This bias is present regardless of the coupling strength, so that the increase of TERV with the coupling strength still can be observed. However, AUROC would not give useful results as the discrimination would be perfect with TERV for all $c$ including $c=0$. The positive bias of the rank measures causes the lack of significance, i.e. obtaining positive values in the absence of coupling, and this has serious implications in real world applications, where also the presence of interaction is investigated.

The TERV measure has an advantage over TE in that it is more stable to noise. For the noise-free coupled R{\"o}ssler-Lorenz system, TE detects clearly the direction of coupling and its performance is enhanced when $T$ increases from 1 to 3. However, when noise is added to the data the estimation of TE is not stable and the large variance does not allow to observe different levels of TE in the two directions. The variance is larger for $T=3$ due to larger dimension of the state space vectors in the estimation of the correlation sums.

We have made the same simulations on two weakly coupled Mackey-Glass systems given as
% ------------------------------------------------------------
\begin{eqnarray}
\frac{\mbox{d}x}{\mbox{d}t} =
\frac{0.2x_{t-\Delta_x}}{1+x_{t-\Delta_x}^{10}}-0.1x_t  \nonumber \\ %
% ---------------------------
\frac{\mbox{d}y}{\mbox{d}t} =
\frac{0.2y_{t-\Delta_y}}{1+y_{t-\Delta_y}^{10}}+ c
\frac{0.2x_{t-\Delta_x}}{1+x_{t-\Delta_x}^{10}}-0.1y_t,
\label{eq:unidMG}
\end{eqnarray}
% ------------------------------------------------------------
where again the driving is from the first system $X$ to the second system $Y$ \cite{Senthilkumar08}. The two systems can have different complexity determined by the delay parameters $\Delta_x$ and $\Delta_y$. We let each $\Delta$ parameter take the values 17, 30, and 100 that, in the absence of coupling, regard systems of correlation dimension at about 2, 3 and 7, respectively \cite{Grassberger83a}. Thus we have 9 different coupled Mackey-Glass systems. All systems are solved using the function {\tt dde23} of the computational environment MATLAB and are sampled at $\tau_s=4$.

The results are quite similar to the results of the R{\"o}ssler-Lorenz system. There is large variation of all measures, and particularly the rank measures, in the detection of direction and strength of coupling, depending on all tested factors: noise, embedding dimensions, future time horizon, and system complexity. The effect of noise is large on TE but small on STE and TERV. Regarding the embedding dimension, $m_x<m_y$ tends to increase the measure with $c$ more in the correct direction $X \rightarrow Y$, $m_x>m_y$ tends to increase the measure in the opposite and false direction $Y \rightarrow X$, and the best balance is obtained for $m_x=m_y$. The rank measures take values at different positive levels in the two directions at no coupling when the two systems have different complexity. Specifically, the direction from the less to more complex system is the one that has the largest positive bias suggesting erroneously causal effect in this direction when $c=0$. This bias may mask the difference of the rank measure values in the two directions for $c>0$. Thus better results can be obtained at small embedding dimensions $m_x=m_y$ and also for $T>1$.

\begin{figure}[htb!]
\centering
\hbox{\includegraphics[height=3.5cm]{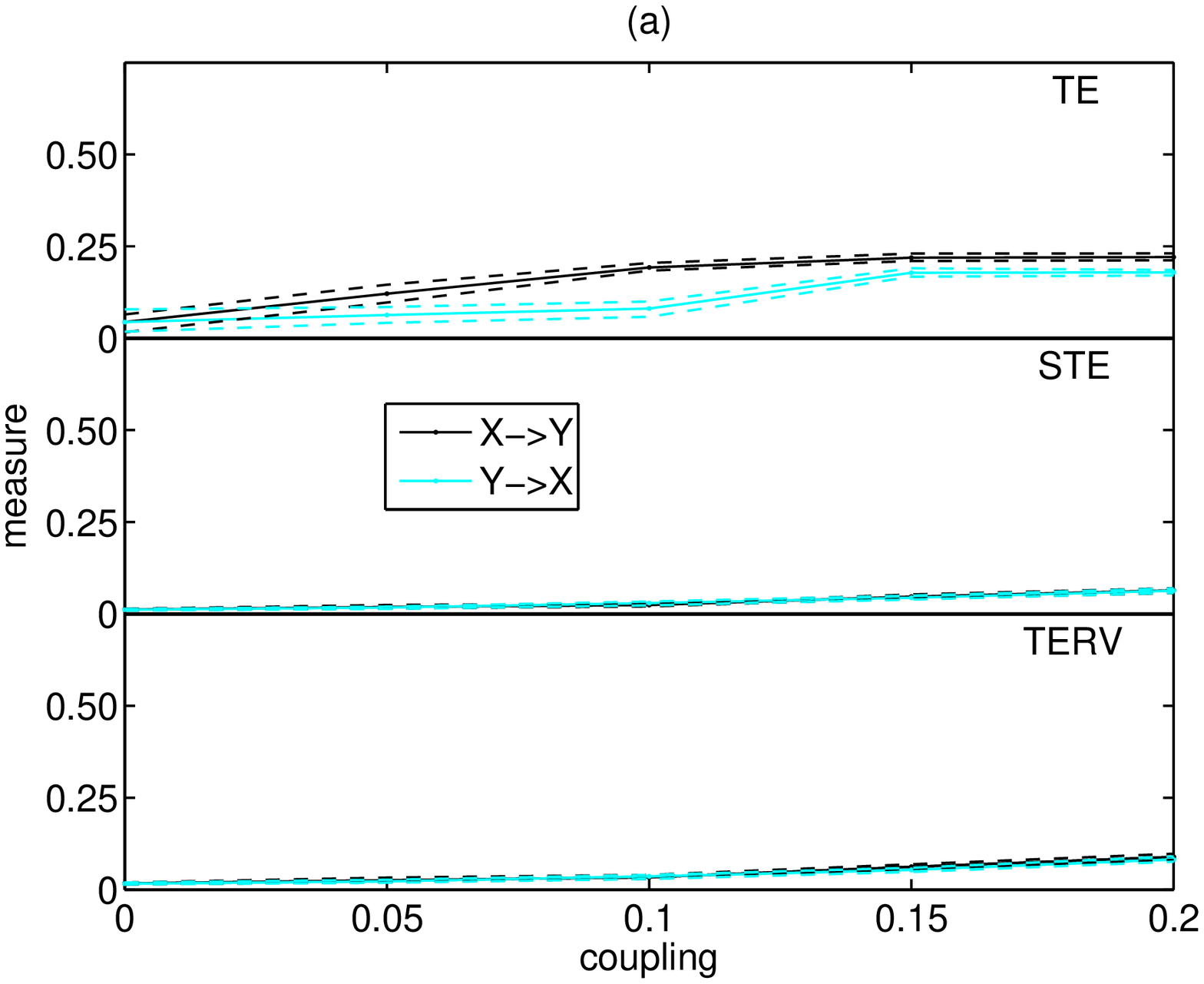}
\includegraphics[height=3.5cm]{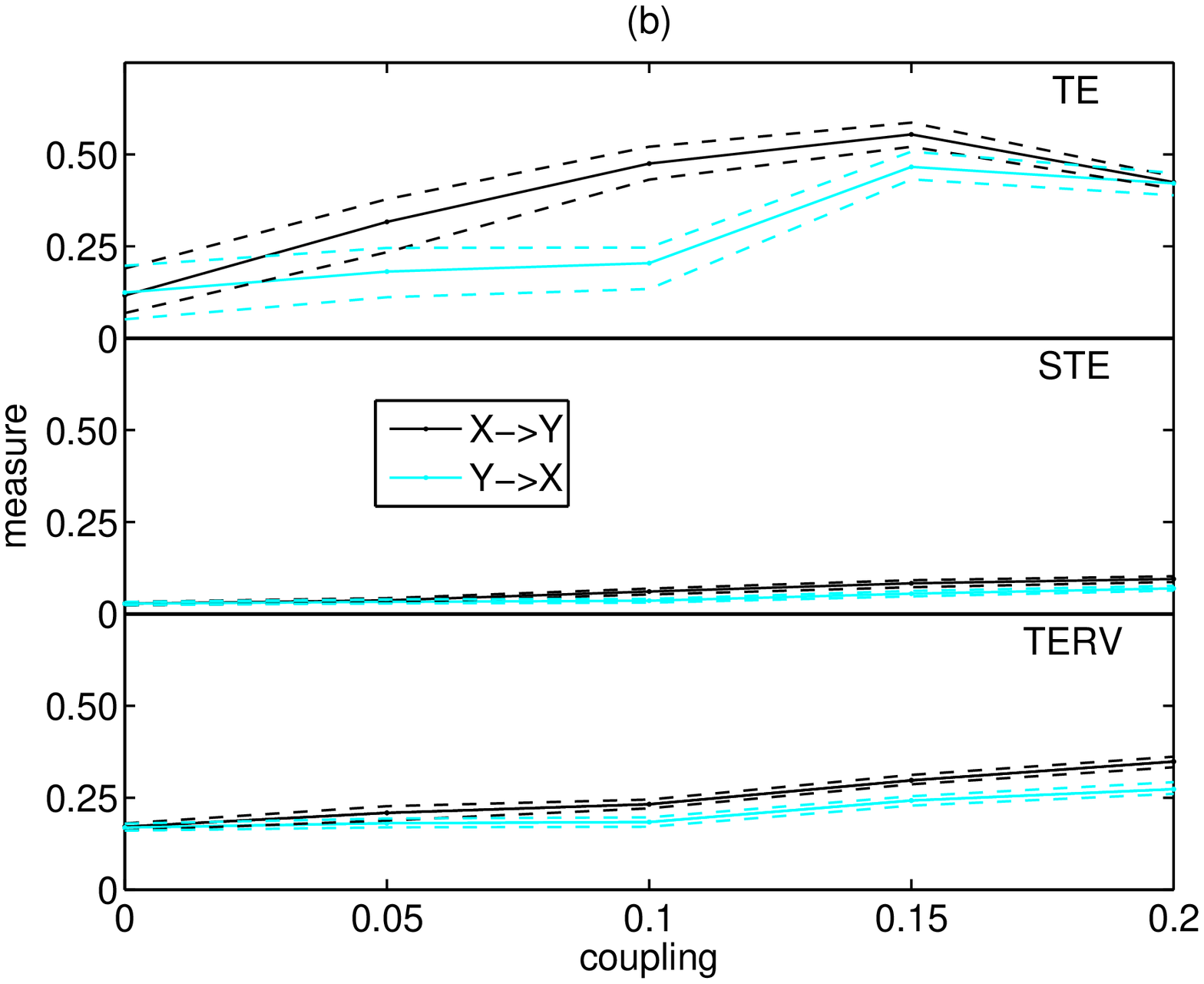}}
\hbox{\includegraphics[height=3.5cm]{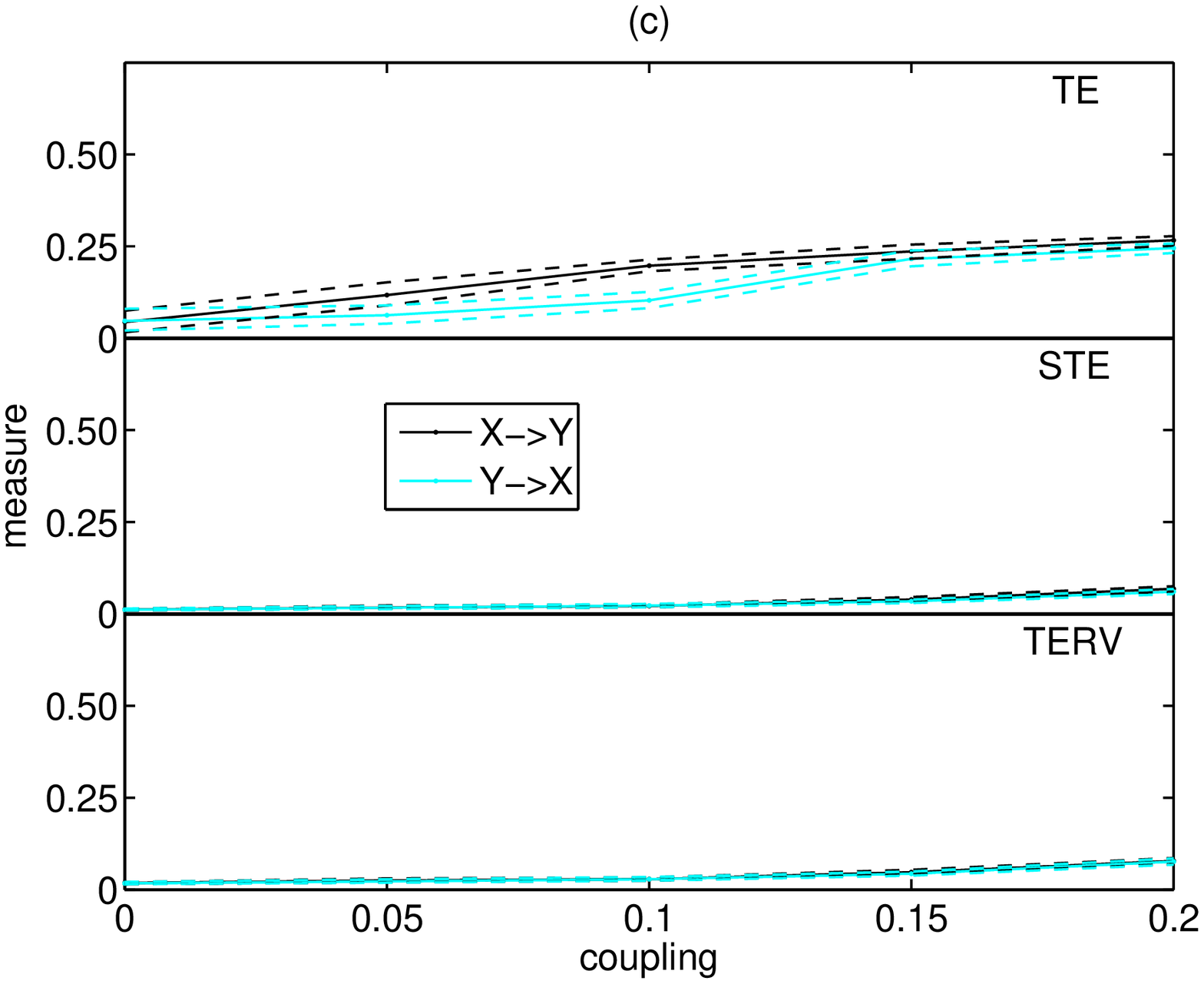}
\includegraphics[height=3.5cm]{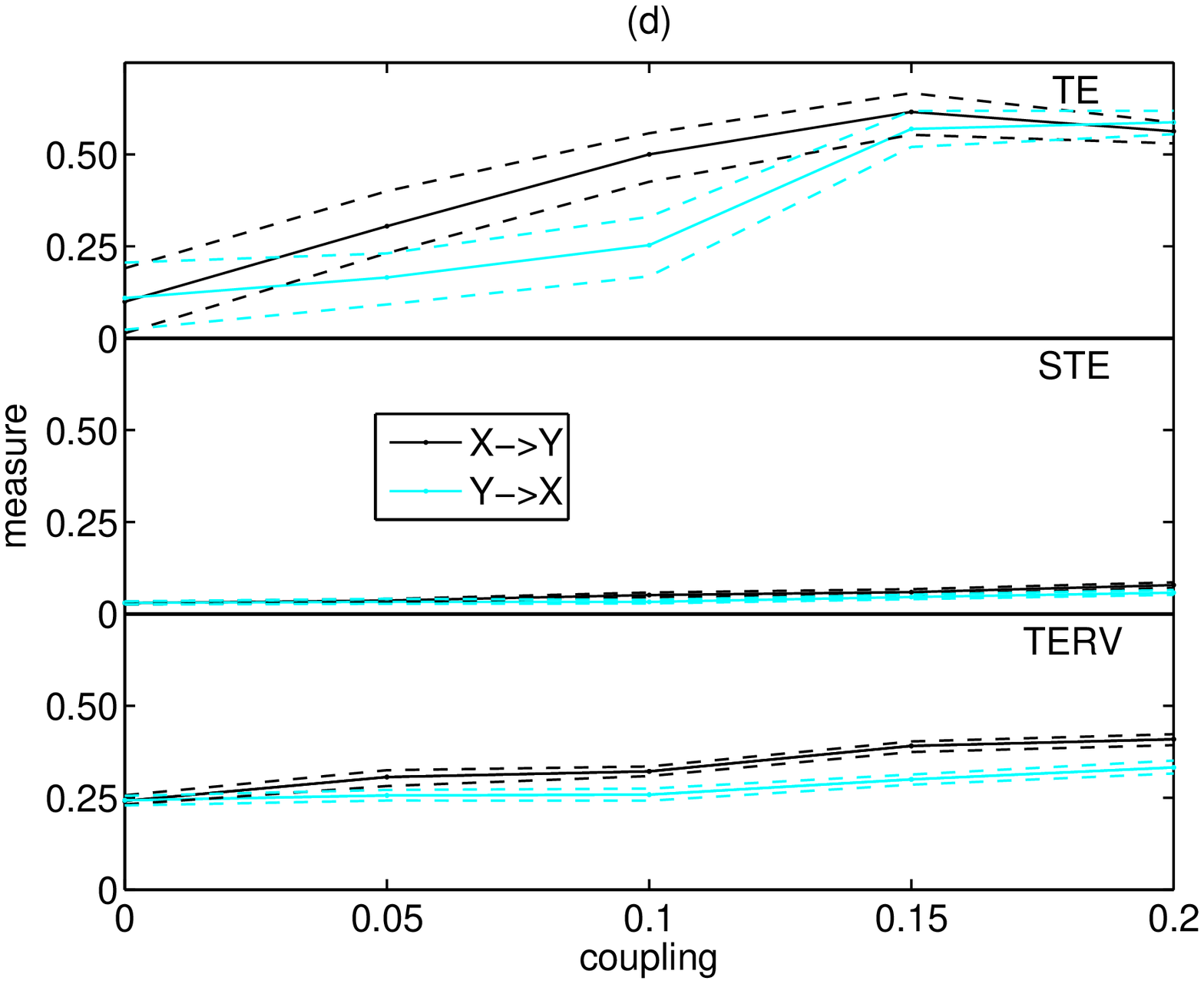}}
\hbox{\includegraphics[height=3.5cm]{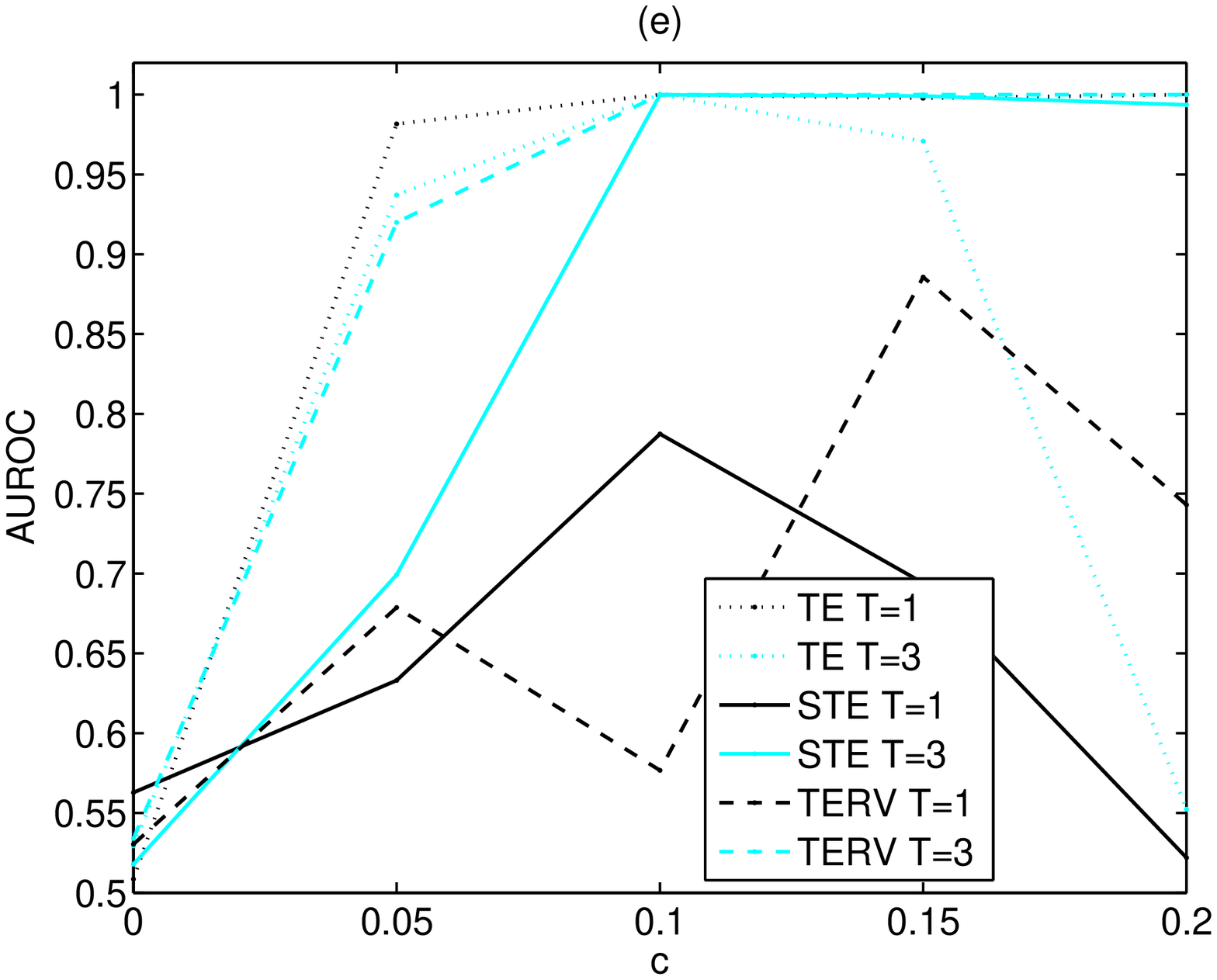}
\includegraphics[height=3.5cm]{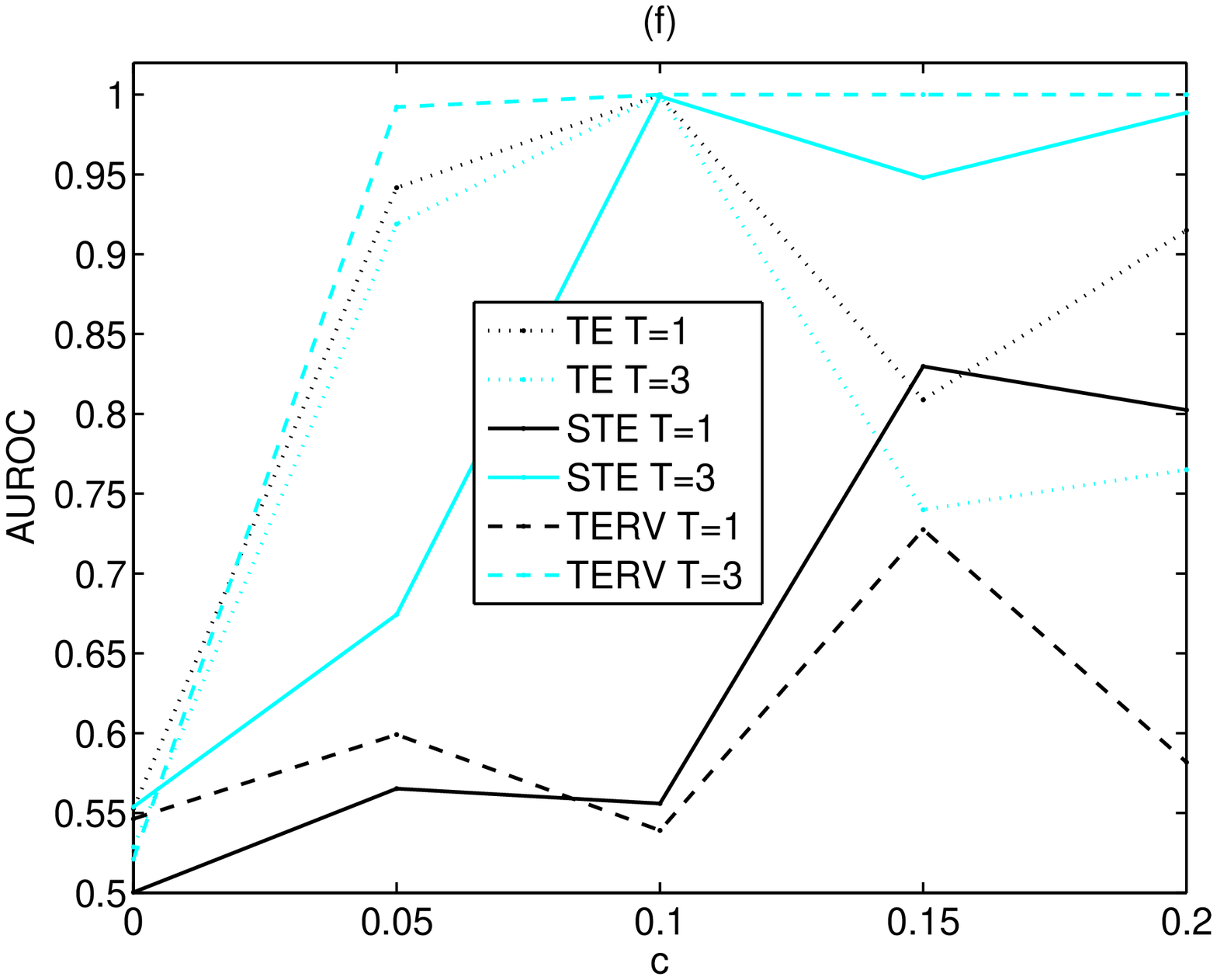}}
\caption{(a-d) As in Fig.~\ref{fig:unidRoeLor}, but for the coupled Mackey-Glass system with $\Delta_x=\Delta_y=30$, $N=4096$ and $m_x=m_y=3$. The noise here is at the level of 10\%. (e) AUROC computed on the measure values displayed in (a) and (b) for the noise-free case and $T=1$ and $T=3$, respectively, as given in the legend. (f) The same as (e) but for the noisy data.}
 \label{fig:unidMG}
\end{figure}

Results for the coupled Mackey-Glass system with $\Delta_x=\Delta_y=30$ and $m_x=m_y=3$ are shown in Fig.~\ref{fig:unidMG}.
For $T=1$ both rank measures do not find differences in the two directions, while TE takes larger values for $c>0$ in the correct direction $X \rightarrow Y$ (see Fig.~\ref{fig:unidMG}a). When $T=3$, both rank measures increase more in the correct direction and give significant differences in the two directions, as shown in Fig.~\ref{fig:unidMG}b. For this example, the rank measures obtain the same level for $c=0$ and AUROC can indeed illustrate the discrimination for $c>0$. As shown in Fig.~\ref{fig:unidMG}e, the AUROC for both rank measures is at the level of the TE measure only for $T=3$. In the presence of noise, TE again tends to have larger variance (even larger for $T=3$) and the discrimination in the two directions is not that clear, particularly for larger coupling strengths ($c=0.15,0.2$), as shown in Fig.~\ref{fig:unidMG}c and d. On the other hand, the rank measures perform similarly to the noise-free case, with TERV performing best and giving the largest AUROC, as shown in Fig.~\ref{fig:unidMG}f.

It should be noted that the overall results on the coupled Mackey-Glass systems are in favor of TE that turns out to be less sensitive than STE and TERV to the variations in system complexity and embedding dimensions, while on the other hand it is more sensitive to the presence of noise.

%%%%%%%%%%%%%%%%%%%%%
\section{Discussion}
\label{sec:discussion}
%%%%%%%%%%%%%%%%%%%%%
The use of ranks of consecutive samples instead of samples themselves in the estimation of the transfer entropy (TE) seems to gain robustness in the presence of noise, a condition often met in real world applications. This was confirmed by our results in the simulation study. Given that TE based on ranks can be a useful measure of information flow and direction of coupling, we have studied the recently proposed rank--based transfer entropy, termed symbolic transfer entropy (STE), and suggested a modified version of STE, which we termed TE on rank vectors (TERV). The first modification is to use the rank of $y_{t+1}$ (one time step ahead for the response time series) in the augmented vector comprised of the reconstructed state vector at time $t$, $\mathbf{y}_t$, and $y_{t+1}$, instead of considering the whole rank vector for $\mathbf{y}_{t+1}$ as done in STE. We showed that indeed this correction gives accurate estimation of the true entropy of the rank vector derived from the joint vector of $\mathbf{y}_{t}$ and $y_{t+1}$. Further, we suggested to allow the time step ahead to be $T>1$ and use the ranks of all samples at the $T$ future times ($y_{t+1},\ldots,y_{t+T}$) derived from the augmented vector containing the current vector $\mathbf{y}_{t}$ and these future samples.

The proposed TERV measure was compared to TE and STE by means of simulations on some known coupled systems, and the level of detection of the coupling direction was also assessed by the area under the receiver operating characteristic curve (AUROC). We found that the detection of the correct coupling direction, as well as the correct identification of uncoupled systems and the estimation of coupling strength when present, varied across the three measures and depended on the presence of noise, the state space reconstruction (we varied both embedding dimensions $m_x$ and $m_y$ for the driver and the response system but used fixed delays $\tau_x=\tau_y=1$), the future horizon $T$, the time series length, and the complexity of the systems. The results are summarized as follows.
\begin{enumerate}
  \item TE estimated by correlation sums has increased variance when noise is added to the data, which may mask the detection of the direction of coupling. STE and TERV are affected much less by noise and often perform better than TE on noisy data.
  \item All measures are dependent on the embedding dimensions and the best results are when equal span of information from the two systems is passed to the reconstructed vectors, i.e. $m_x=m_y$, a condition set arbitrarily, but apparently correctly as our simulations justify, in most works on coupling measures. If the embedding dimension for the driving system is larger the measure tends to be larger in the wrong direction of interaction. Rank measures (STE and TERV) tend to be more sensitive to the selection of the two embedding dimensions than TE.
  \item When information flow is measured by TE and TERV over a future horizon of length $T>1$ it can detect better than for $T=1$ the correct direction and strength of coupling, provided that the estimation of the entropy terms is stable. Note that using a larger $T$ increases the dimension of the future response vector in the definition of TE and TERV and consequently the data requirements. Thus the stability of the estimation depends on the length of the time series, the level of noise in the data and the two embedding dimensions. STE does not show not the same improvement in performance when $T>1$ due to the way the rank future vector of the response is constructed.
  \item The measures using ranks (STE and TERV) have larger positive bias than TE that depends on embedding dimension, time series length and system complexity. For example, all measures increase with the embedding dimension (also when $m_x=m_y$), so that even in the absence of coupling there are significantly larger than zero. In the simulations with increasing coupling strength $c$, the difference of the measures on uncoupled and coupled systems still could be maintained, but in applications where a single case of coupling is to be investigated, the lack of significance.
  \item For different complexity there is different bias in the two directions and the rank measures tend to differ at $c=0$. TE shows this effect at a lesser extent. The largest bias is in the direction from the less to more complex system. When the two systems are of the same complexity, the bias is the same in both directions allowing the rank measures to detect well the coupling direction and the strength of coupling.
\end{enumerate}
Given the above finding, overall TERV gave better discrimination of the direction of coupling (higher AUROC) than STE, and when the data were noisy also better than TE in many cases. In particular, the use of $T>1$ improved the performance of TERV and TE but not STE.

The results on TE yield the particular estimate using correlation sums. A small scale simulation has showed that binning estimates performed worse, especially when the dimension increased (embedding dimension and $T$), and this is attributed to the problem of binning for high state space dimensions. On the other hand, the nearest neighbor estimate \cite{Kraskov04} was more stable, particularly on noisy data and high dimensions. Further investigation on the estimates of TE is obviously needed.

\footnotesize

% Bibliography----------------------------------------------------------------


\begin{thebibliography}{1}

\bibitem{Granger69}
J.~Granger, ``Investigating causal relations by econometric models and
  cross-spectral methods,'' \emph{Acta Physica Polonica B}, vol.~37, pp. 424
  –- 438, 1969.

\bibitem{Baccala01}
L.~Baccala and K.~Sameshima, ``Partial directed coherence: a new concept in
  neural structure determination,'' \emph{Biological Cybernetics}, vol.~84,
  no.~6, pp. 463 -- 474, 2001.

\bibitem{Winterhalder05}
M.~Winterhalder, B.~Schelter, W.~Hesse, K.~Schwab, L.~Leistritz, D.~Klan,
  R.~Bauer, J.~Timmer, and H.~Witte, ``Comparison of linear signal processing
  techniques to infer directed interactions in multivariate neural systems,''
  \emph{Signal Processing}, vol.~85, no.~11, pp. 2137 -- 2160, 2005.

\bibitem{Rosenblum01}
M.~Rosenblum and A.~Pikovski, ``Detecting direction of coupling in interacting
  oscillators,'' \emph{Physical Review E}, vol.~64, no.~4, Article 045202, 2001.

\bibitem{QuianQuiroga02}
R.~Quian~Quiroga, T.~Kreuz, and P.~Grassberger, ``Event synchronization: A
  simple and fast method to measure synchronicity and time delay patterns,''
  \emph{Physical Review E}, vol.~66, no.~4, Article 041904, 2002.

\bibitem{Smirnov03}
D.~Smirnov and B.~Bezruchko, ``Estimation of interaction strength and direction
  from short and noisy time series,'' \emph{Physical Review E}, vol.~68, no.~4,
  Article 046209, 2003.

\bibitem{Cenys92}
A.~Cenys, G.~Lasiene, K.~Pyragas, J.~Peinke, and J.~Parisi, ``Analysis of
  spatial correlations in chaotic systems,'' \emph{Acta Physica Polonica B},
  vol.~23, no.~4, pp. 357 -- 365, 1992.

\bibitem{Schiff96}
S.~Schiff, P.~So, T.~Chang, R.~Burke, and T.~Sauer, ``Detecting dynamical
  interdependence and generalized synchrony through mutual prediction in a
  neural ensemble,'' \emph{Physical Review E}, vol.~54, pp. 6708 -- 6724, 1996.

\bibitem{Arnhold99}
J.~Arnhold, P.~Grassberger, K.~Lehnertz, and C.~Elger, ``A robust method for
  detecting interdependences: Application to intracranially recorded {EEG},''
  \emph{Physica D}, vol. 134, pp. 419 -- 430, 1999.

\bibitem{QuianQuiroga00b}
R.~Quian~Quiroga, J.~Arnhold, and P.~Grassberger, ``Learning driver-response
  relationships from synchronization patterns,'' \emph{Physical Review E},
  vol.~61, no.~5, pp. 5142 -- 5148, 2000b.

\bibitem{Andrzejak03b}
R.G.~Andrzejak, A.~Kraskov, H.~St{\"{o}}gbauer, F.~Mormann, and T.~Kreuz,
  ``Bivariate surrogate techniques: Necessity, strengths, and caveats,''
  \emph{Physical Review E}, vol.~68, Article 066202, 2003.

\bibitem{Romano07}
M.~Romano, M.~Thiel, J.~Kurths, and G.~C., ``Estimation of the direction of the
  coupling by conditional probabilities of recurrence,'' \emph{Physical Review
  E}, vol.~76, no.~3, Article 036211, 2007.

\bibitem{Chicharro09}
D.~Chicharro and R.G.~Andrzejak,
``Reliable Detection of Directional Couplings Using Rank Statistics,''
\emph{Physical Review E}, vol.~80, Article 026217, 2009.

\bibitem{Schreiber00}
T.~Schreiber, ``Measuring information transfer,'' \emph{Physical Review
  Letters}, vol.~85, no.~2, pp. 461 -- 464, 2000.

\bibitem{Palus01}
M.~Palu\v{s}, V.~Kom\'{a}rek, T.~Proch\'{a}zka, Z.~Hrnc\'{i}r, and
  K.~\v{S}terbov\'{a}, ``Synchronization and information flow in {EEG}s of
  epileptic patients,'' \emph{IEEE Engineering in Medicine and Biology
  Magazine}, vol.~20, no.~5, pp. 65--71, 2001.

\bibitem{Marschinski02}
R.~Marschinski and H.~Kantz, ``Analysing the information flow between financial
  time series,'' \emph{European Physical Journal B}, vol.~30, pp. 275 -- 281,
  2002.

\bibitem{Staniek08}
M.~Staniek and K.~Lehnertz, ``Symbolic transfer entropy,'' \emph{Physical
  Review Letters}, vol. 100, no.~15, Article 158101, 2008.

\bibitem{Vejmelka08}
M.~Vejmelka and M.~Palu\v{s}, ``Inferring the directionality of coupling with
  conditional mutual information,'' \emph{Physical Review E}, vol.~77, no.~2,
  Article 026214, 2008.

\bibitem{Bahraminasab08}
A.~Bahraminasab, F.~Ghasemi, A.~Stefanovska, P.~V.~E. McClintock, and H.~Kantz,
  ``Direction of coupling from phases of interacting oscillators: A permutation
  information approach,'' \emph{Physical Review Letters}, vol. 100, no.~8, Article
  084101, 2008.

\bibitem{Lungarella07}
M.~Lungarella, K.~Ishiguro, Y.~Kuniyoshi, and N.~Otsu, ``Methods for
  quantifying the causal structure of bivariate time series,'' \emph{Journal of
  Bifurcation and Chaos}, vol.~17, no.~3, pp. 903 -- 921, 2007.

\bibitem{Palus07}
M.~Palu\v{s} and M.~Vejmelka, ``Directionality of coupling from bivariate time
  series: How to avoid false causalities and missed connections,''
  \emph{Physical Review E}, vol.~75, no.~5, Article 056211, 2007.

\bibitem{Papana08b}
A.~Papana and D.~Kugiumtzis, ``Detection of directionality of information
  transfer in nonlinear dynamical systems,'',
\emph{Topics on Chaotic Systems, selected papers from {CHAOS} 2008 International Conference},
World Scientific, pp. 251 -- 264, 2009.

\bibitem{Sabesan09b}
S.~Sabesan, L.~B.~Good, K.~S.~Tsakalis, A.~Spanias, D.~M.~Treiman and L.~D.~Iasemidis,
``Information Flow and Application to Epileptogenic Focus Localization from Intracranial EEG'',
\emph{IEEE Transactions on Neural Systems and Rehabilitation Engineering}, vol.~17, no. 3, pp. 244 -- 253, 2009.

\bibitem{Kwon08}
O.~Kwon and J.-S.~Yang,
``Information Flow between Stock Indices'',
\emph{EPL (Europhysics Letters)}, vol.~82, no. 6, Article 68003, 2008.

\bibitem{Cover91}
T.~Cover and J.~Thomas, \emph{Elements of Information Theory}.\hskip 1em plus
  0.5em minus 0.4em\relax New York: John Wiley and Sons, 1991.

\bibitem{Papana09}
A.~Papana and D.~Kugiumtzis, ``Evaluation of Mutual Information Estimators for Time Series,'' \emph{International Journal of Bifurcation and Chaos}, vol.~19, no.~12, pp. 4197 -- 4215, 2009.
  
\bibitem{Silverman86}
B.~Silverman, \emph{Density Estimation for Statistics and Data Analysis}.\hskip
  1em plus 0.5em minus 0.4em\relax London: Chapman and Hall, 1986.

\bibitem{Kraskov04}
A.~Kraskov, H.~St{\"{o}}gbauer, and P.~Grassberger, ``Estimating mutual
  information,'' \emph{Physical Review E}, vol.~69, no.~6, Article 066138, 2004.

\bibitem{Diks02}
C.~Diks and S.~Manzan, ``Tests for Serial Independence and Linearity Based on
  Correlation Integrals,'' \emph{Studies in Nonlinear Dynamics \&
  Econometrics}, vol.~6, no. 2, Article 2, 2002.

\bibitem{Hand01}
D.~J. Hand and R.~J. Till, ``A simple generalization of the area under the
  {ROC} curve to multiple class classification problems,'' \emph{Machine
  Learning}, vol.~45, pp. 171 -- 186, 2001.

\bibitem{Kugiumtzis09}
D.~Kugiumtzis, ``Improvement of Symbolic Transfer Entropy'',
\emph{3rd International Conference on Complex Systems and Applications, Conference Proceedings, Special Sessions},
Eds C.~Bertelle, X.~Liu and M.~A. Aziz-Alaoui, pp. 338 -- 342, 2009.

\bibitem{Quyen99}
M.~Le~Van~Quyen, J.~Martinerie, C.~Adam and F.~J.~Varela,
``Nonlinear Analyses of Interictal {EEG} Map the Brain Interdependences in Human Focal Epilepsy'',
\emph{Physica D: Nonlinear Phenomena}, vol.~127, no~3-4, pp. 250 -- 266, 1999.

\bibitem{Senthilkumar08}
D.~V.~Senthilkumar, M.~Lakshmanan and J.~Kurths,
``Transition from Phase to Generalized Synchronization in Time-Delay Systems'',
\emph{Chaos: An Interdisciplinary Journal of Nonlinear Science}, vol.~18, no~2, Article 023118, 2008.

\bibitem{Grassberger83a}
P.~Grassberger and I.~Procaccia,
``Measuring the Strangeness of Strange Attractors'',
\emph{Physica D: Nonlinear Phenomena}, vol.~9, pp. 189 -- 208, 1983.

\end{thebibliography}
\end{document}